\documentclass[aps,prl,reprint,superscriptaddress,showpacs]{revtex4-1}

\usepackage{graphicx}

\begin{document}


\title{Magnetic Moment Formation in Graphene Detected by Scattering of Pure Spin Currents}


\author{Kathleen M. McCreary}
\thanks{These authors contributed equally.}
\affiliation{Department of Physics and Astronomy, University of California, Riverside,
CA 92521, USA}

\author{Adrian G. Swartz}
\thanks{These authors contributed equally.}
\affiliation{Department of Physics and Astronomy, University of California, Riverside,
CA 92521, USA}

\author{Wei Han}
\affiliation{Department of Physics and Astronomy, University of California, Riverside,
CA 92521, USA}

\author{Jaroslav Fabian}
\affiliation{Institute for Theoretical Physics, University of Regensburg, D-93040 Regensburg, Germany}

\author{Roland K. Kawakami}
\email[]{roland.kawakami@ucr.edu}
\affiliation{Department of Physics and Astronomy, University of California, Riverside,
CA 92521, USA}


\date{\today}

\begin{abstract}
Hydrogen adatoms are shown to generate magnetic moments inside single layer graphene. Spin transport measurements on graphene spin valves exhibit a dip in the non-local spin signal as a function of applied magnetic field, which is due to scattering (relaxation) of pure spin currents by exchange coupling to the magnetic moments. Furthermore, Hanle spin precession measurements indicate the presence of an exchange field generated by the magnetic moments. The entire experiment including spin transport is performed in an ultrahigh vacuum chamber, and the characteristic signatures of magnetic moment formation appear only after hydrogen adatoms are introduced. Lattice vacancies also demonstrate similar behavior indicating that the magnetic moment formation originates from {\it{p$_z$}}-orbital defects.
\end{abstract}


\maketitle


Many fascinating predictions have been made regarding magnetism in graphene including the formation of magnetic moments from dopants, defects, and edges \cite{yazyev:2007,
boukhvalov:2008,zhou:2009,soriano:2011,uchoa:2008,vozmediano:2005,
pisani:2007,lee:2005}. While several experimental techniques provide insight into this problem 
\cite{nair:2012,yazyev:2010,sepioni:2010,wang:2009,ramakrishna:2009,
xie:2011,esquinazi:2003,cervenka:2009,candini:2011,chen:2011,hong:2011,
ugeda:2010,tao:2011}, lack 
of clear evidence for magnetic moment formation hinders development of this nascent field.  Studies based on bulk magnetometry \cite{nair:2012,yazyev:2010,sepioni:2010,wang:2009,ramakrishna:2009,xie:2011,esquinazi:2003,cervenka:2009} directly measure magnetic properties, but because it measures the total magnetic moment (not just the signal from graphene) it is difficult to rule out artifacts from environmental magnetic impurities. Transport \cite{candini:2011,chen:2011,hong:2011}
and scanning tunneling microscopy (STM)  \cite{ugeda:2010,tao:2011} locally probe the graphene, but so far these measurements have been charge-based, so data are subject to various interpretations \cite{jobst:2012}. Thus, in order to convincingly demonstrate the formation of magnetic moments inside graphene due to dopants and defects, it is essential to employ techniques that directly probe the intrinsic spin degree-of-freedom of the magnetic moment while ensuring the signal originates from the graphene sheet under investigation.

In this Letter, we utilize pure spin currents to demonstrate that hydrogen adatoms and lattice vacancies generate magnetic moments in single layer graphene. Pure spin currents are injected into graphene spin valve devices and clear signatures of magnetic moment formation emerge in the non-local spin transport signal as hydrogen adatoms or lattice vacancies are systematically introduced in an ultrahigh vacuum (UHV) environment. Specifically, introduction of these point defects generate a characteristic dip in the non-local signal as a function of magnetic field. This feature is due to scattering (relaxation) of pure spin currents by localized magnetic moments in graphene and is explained quantitatively by a phenomenological theory based on spin-spin exchange coupling between conduction electrons and magnetic moments. Furthermore, we observe effective exchange fields due to this spin-spin coupling, which are of interest for novel phenomena and spintronic functionality \cite{haugen:2008,semenov:2007,michetti:2010,qiao:2010} but have not been seen previously in graphene. Thus, these results provide the most clear and direct evidence for magnetic moment formation in graphene and demonstrate a method for utilizing localized magnetic moments to manipulate conduction electron spins. 	

For a systematic investigation, the spin transport measurement is first performed on a pristine single layer graphene (SLG) spin valve as a control measurement. Then, dopants/defects are controllably introduced to the SLG and the measurement is repeated. The sample remains in UHV during the entire process. Therefore, observed signatures of magnetic moment formation are caused by the adsorbed hydrogen or lattice vacancies. 

\begin{figure}
\includegraphics[width=90mm]{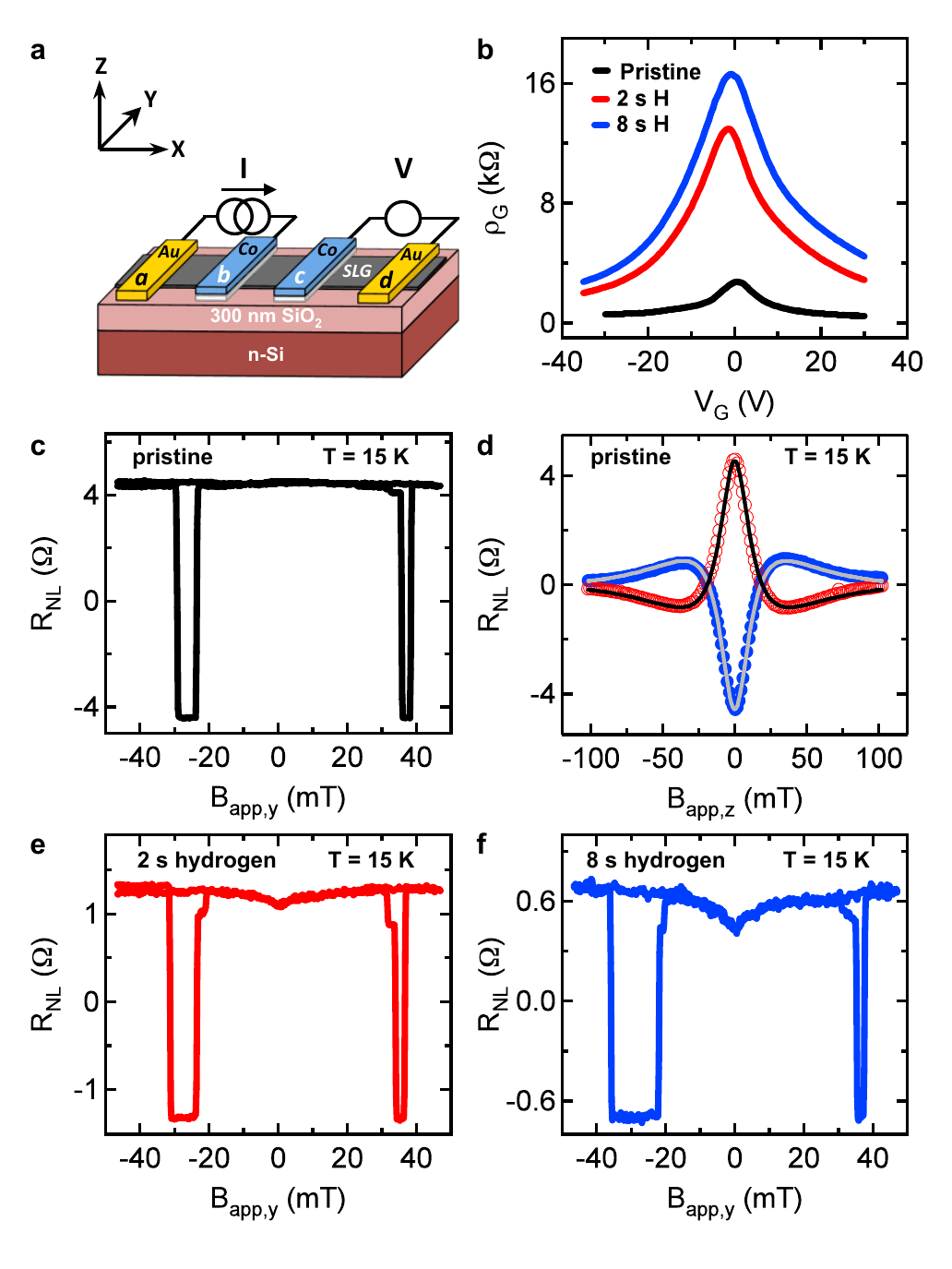}
\caption{\label{fig:main}The effect of hydrogen exposure on charge and spin transport in SLG at 15 K. (a) Schematic illustration of the non-local spin valve device. (b) Gate dependent resistivity for the pristine graphene (black) and following exposure to atomic hydrogen for 2 s (red) and 8 s (blue). Upon hydrogen doping, the Dirac point shifts from 0 V to -1 V. (c) Non-local spin transport  measurement for pristine graphene. (d) Hanle spin precession measurement on pristine graphene. (e,f) Non-local spin transport measurements after atomic hydrogen exposure for 2 s and 8 s, respectively. Both curves exhibit a dip in $R_{NL}$ at zero applied field, which is caused by spin relaxation induced by localized magnetic moments.}
\end{figure}

Experiments are performed on non-local SLG spin valves
\cite{johnson:1985,tombros:2007,han:2012} (Fig.~\ref{fig:main}a) consisting of two outer Au/Ti electrodes (a and d) and two ferromagnetic (FM) Co electrodes that make contact to SLG across MgO/TiO$_2$ tunnel barriers (b and c). The Co electrodes are capped with 5 nm Al$_2$O$_3$ to protect from hydrogen exposure.  The tunnel barrier and capping layer are present only at the site of the FM electrodes, leaving the rest of the graphene uncovered. The device is fabricated on a SiO$_2$/Si substrate (300 nm thickness of SiO$_2$) where the Si is used as a back gate. Details of device fabrication are published elsewhere \cite{han:2012}.

The charge and spin transport properties of pristine SLG spin valves are measured at 15 K using lock-in techniques. The gate dependent resistivity ($\rho_G$) of a representative sample A (black curve in Fig.~\ref{fig:main}b) exhibits a maximum at the gate voltage ($V_G$) of 0 V, which defines the Dirac point ($V_D =$ 0 V). This sample exhibits mobility ($\mu$) of 6105 cm$^2$/Vs. To investigate spin transport in the SLG device (Fig.~\ref{fig:main}a), a current ($I$) is applied between electrodes b and a, injecting spin-polarized carriers into graphene directly below the FM injector, b. The spin population diffuses along the sample as a pure spin current ($x$-axis) and the spin density is measured at the FM spin detector, c, as a voltage difference ($V$) between electrodes c and d. An applied magnetic field ($B_{app,y}$) along the electrode magnetization direction ($y$-axis) is used to control the relative orientation of spin injector and detector magnetizations. For parallel alignment, the measured non-local resistance ($R_{NL}=V/I$) is positive whereas for antiparallel alignment $R_{NL}$ is negative. The non-local spin signal is defined as the difference between parallel and antiparallel states ($\Delta R_{NL}=R^P_{NL}-R^{AP}_{NL}$ ). A typical scan of $R_{NL}$ as a function of $B_{app,y}$ (Fig.~\ref{fig:main}c) displays discrete jumps as the electrode orientation changes between parallel and antiparallel. 
This sample exhibits a $\Delta R_{NL}$ 
of 8.8 $\Omega$ (sample A with $V_G - V_D$ = -15 V). A constant spin-independent background is subtracted from all $R_{NL}$ data presented in this study. Out-of-plane magnetic fields are applied to generate spin precession, and the resulting data (Fig.~\ref{fig:main}d, red for parallel, blue for antiparallel) are fit by the standard Hanle equation \cite{tombros:2007,han:2012} (solid curves) to determine the spin lifetime ($\tau_{so}$ = 479 ps) and diffusion coefficient ($D$ = 0.023 m$^2$/s ). The corresponding spin diffusion length is  $\lambda = \sqrt{D \tau^{so}}$ = 3.3 $\mu$m. Based on these values and a non-local spin signal of 8.8 $\Omega$, the spin polarization of the junction current ($P_J$) is calculated to be 20\% \cite{takahashi:2003,supplement}.

Atomic hydrogen is introduced to spin valve devices at 15 K at a chamber pressure of $1\times 10^{-6}$ torr  \cite{supplement}. Following 2 s hydrogen exposure, the gate dependent $\rho_G$ (red curve in Fig.~\ref{fig:main}b) is dramatically increased. An additional 6 s of exposure (8 s total) further increases $\rho_G$ (blue curve of Fig.~\ref{fig:main}b) and decreases the mobility to 495 cm$^2$/Vs. Based on the change in the resistivity, we make an order of magnitude estimate for the hydrogen coverage of 0.1\% \cite{supplement}. Accompanying the changes in charge transport are also changes in spin transport. Figures 1e and 1f display 
$R_{NL}$ of sample A at $V_G - V_D$ = -15 V as a function of $B_{app,y}$ following 2 s and 8 s of exposure, respectively. The initial $\Delta R_{NL}$ of 8.8 $\Omega$ is reduced to 2.6 $\Omega$
after 2 s of hydrogen exposure and further reduced to 1.4 $\Omega$ after 8 s. 
Interestingly, the $R_{NL}$ scans exhibit a dip centered at zero applied field. The dip in $R_{NL}$ is prevalent for both up and down sweeps of $B_{app,y}$ at all measured gate voltages and has been reproduced on multiple samples following hydrogen exposure. The ratio of the dip magnitude to $\Delta R_{NL}$  is found to increase with increasing hydrogen exposure (comparing Fig.~\ref{fig:main}e and \ref{fig:main}f), indicating the dip feature is dependent on the amount of adsorbed hydrogen. 

\begin{figure}
\includegraphics[width=85mm]{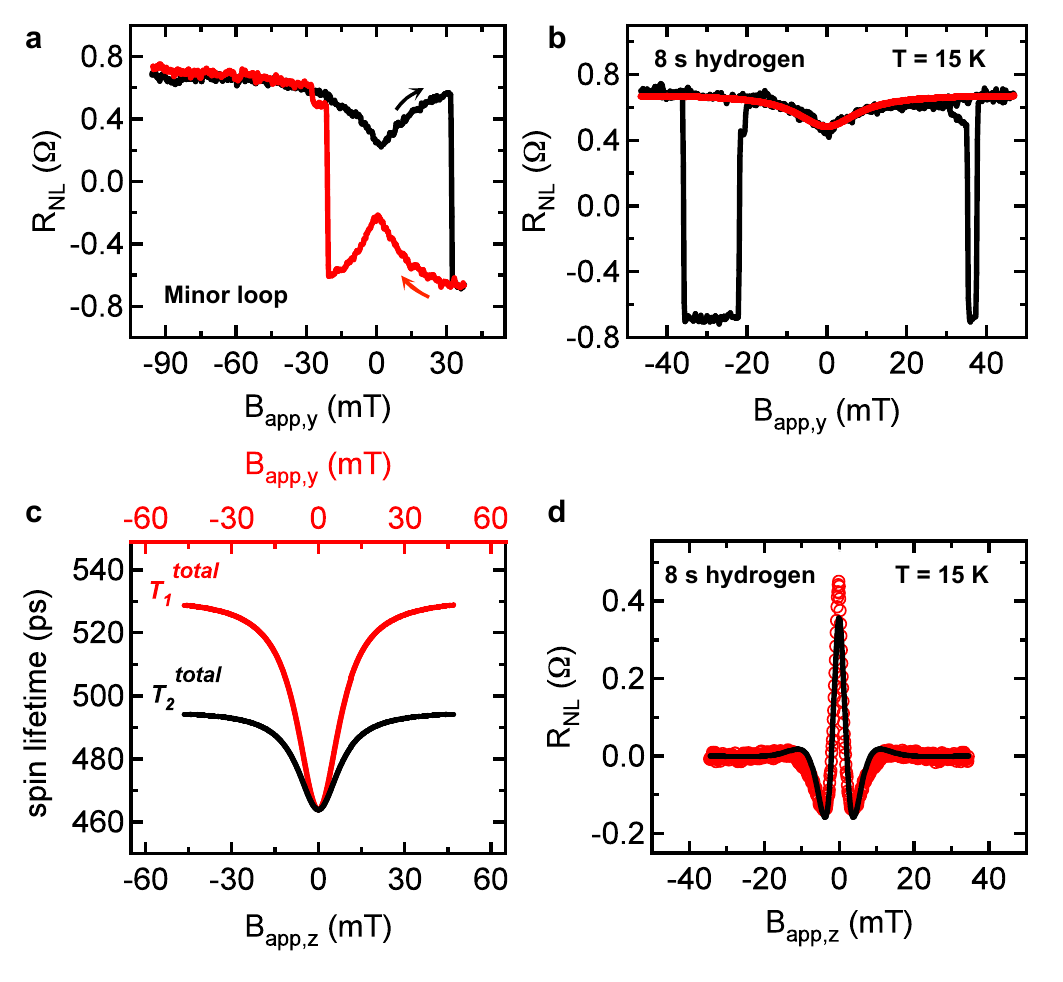}
\caption{\label{fig:analysis} (a) A minor loop scan shows that the dip in $R_{NL}$ for parallel alignment (black) becomes a peak for antiparallel alignment (red), indicating the feature is due to increased spin relaxation, as opposed to an artifact of the background level. (b) Fitting the dip in $R_{NL}$ based on the model of spin relaxation by paramagnetic moments (data in black, fit in red). (c) Field dependence of longitudinal (red) and transverse (black) spin lifetimes. (d) Hanle precession data following 8 s hydrogen exposure (red) is fit using equation \ref{eq:hanle} (black curve).
}
\end{figure}

To understand the origin of the dip in $R_{NL}$, we examine the expression for non-local resistance generated by spin transport \cite{takahashi:2003},
\begin{eqnarray}
R^{(P/AP)}_{NL}=\pm 2R_Ge^{-L/\lambda} \prod^{2}_{i=1}
\left( \frac{P_J \frac{R_i}{R_G}}{1-P_J^2} +  \frac{P_F \frac{R_F}{R_G}}{1-P_F^2}
\right)
\nonumber\\
 \times
\left[\prod^{2}_{i=1}
\left( 1+ \frac{2 \frac{R_i}{R_G}}{1-P_J^2} + \frac{2 \frac{R_F}{R_G}}{1-P_F^2}
\right) - e^{-2L/\lambda}
\right]^{-1}
\label{eq:nonlocal}
\end{eqnarray}
where $R_G =\rho_G \lambda/w$ is the spin resistance of graphene, $w$ is the graphene width, $R_F = \rho_F \lambda_F/A_J$ is the spin resistance of the cobalt, $\rho_{F}$ is the cobalt resistivity, $\lambda_{F}$ is the cobalt spin diffusion length, $A_J$ is the junction area, $P_F$ is the spin polarization of cobalt, $R_1$ and $R_2$ are the contact resistances of the spin injector and detector, respectively, and $L$ is the distance from injector to detector. This equation shows that the spin density at the detector electrode depends on both charge and spin properties. First, we confirm that the SLG resistivity does not change with magnetic field, so the dip is not related to changes in charge transport  \cite{supplement}. Second, we verify that the dip is not related to hydrogen-induced changes to the magnetic properties of the FM electrodes. Specifically, the effect of hydrogen exposure is reversible upon thermal cycling to room temperature and the anisotropic magnetoresistance of the Co electrodes are not affected by hydrogen exposure \cite{supplement}. Next, we perform minor loop analysis on sample B (Fig.~\ref{fig:analysis}a) by reversing the magnetic field sweep immediately after the first magnetization reversal. The inversion of the dip in the antiparallel state (red curve) proves that the dip is due to increased spin relaxation at low fields. Furthermore, we rule out hyperfine coupling to nuclear spins as the origin of this increased 
spin relaxation  \cite{supplement}.

As we discuss in the following, emergence of the dip following hydrogen adsorption identifies magnetic moment formation in graphene. The dip in $R_{NL}$ is a characteristic feature of spin relaxation from exchange coupling with localized magnetic moments, and can be illustrated from a simple textbook example of two coupled spins in a magnetic field. The Hamiltonian is given by  
$H=A_{ex} \vec{S}_e \cdot \vec{S}_M  + g_e \mu_B \vec{S}_e \cdot \vec{B}_{app} + g_M \mu_B \vec{S}_M \cdot \vec{B}_{app}$,
where $\vec{S}_e$   is the conduction electron spin,   $\vec{S}_M$ is the spin of the magnetic moment, $g_e$ and $g_M$ are the respective $g$-factors, and $A_{ex}$ is the exchange coupling strength \cite{furdyna:1988, bastard:1991}. Due to the presence of the exchange coupling, the individual spins are not conserved; only the total spin $\vec{S}_{tot}=\vec{S}_e+\vec{S}_M$  is conserved. For the case where both 
$\vec{S}_e$  and $\vec{S}_M$  are spin-$\frac{1}{2}$, the quantum mechanical eigenstates in zero magnetic field are the well-known singlet ($S_{tot}=0$) and triplet ($S_{tot}=1$) spin states \cite{griffiths}. At higher magnetic fields the Zeeman terms dominate and the two spins decouple so that the magnitudes and $z$-components of $\vec{S}_e$ and $\vec{S}_M$ become good quantum numbers, similar to the Paschen-Back effect \cite{griffiths}. Thus, the dip in $R_{NL}$ is qualitatively explained by the non-conservation of $\vec{S}_e$ at low fields due to the presence of exchange coupling with magnetic moments.

To quantitatively analyze the experimental data, we must consider that a conduction electron will interact with many localized magnetic moments. Thus, the terms in the Hamiltonian involving the conduction electron are given by
$H_e = \eta_M A_{ex} \vec{S}_e \cdot \langle \vec{S}_M \rangle+g_e \mu_B \vec{S}_e \cdot \vec{B}_{app} = g_e \mu_B \vec{S}_e \cdot \left( \overline{\vec{B}}_{ex}+ \vec{B}_{app} \right)$
where $\eta_M$ is the filling density of magnetic moments. The averaging $\langle \cdots \rangle$  is over the ensemble of magnetic moments and the effective field generated by the exchange interaction is $\overline{\vec{B}}_{ex}=\frac{\eta_M A_{ex} \langle \vec{S}_M \rangle }{g_e \mu_B}$. As the spins diffuse through the lattice they experience varying magnetic moments which results in varying Larmor frequencies. In the local frame associated with the electrons this can be described by a time-dependent, randomly fluctuating magnetic field, 
$\vec{B}_{ex}(t)=\overline{\vec{B}}_{ex}+\Delta \vec{B}_{ex}(t)$.
For the  $R_{NL}$ measurements, the longitudinal spin relaxation due to a fluctuating field is 
given by \cite{fabian:2007}, 
\begin{equation}
\frac{1}{\tau^{ex}_1} =\frac{ \left( \Delta B \right)^2}{\tau_c}
\frac{1}{\left( B_{app,y}+ \overline{B}_{ex,y} \right)^2  + \left( \frac{\hbar}{g_e \mu_B \tau_c } \right)^2 }
\label{eq:relax}
\end{equation}								
where $\Delta B$ is the rms fluctuation and $\tau_c$ is the correlation time  \cite{supplement}. 
The spin relaxation rate due to the exchange field is described by a Lorentzian curve which depends explicitly on the applied field, $B_{app,y}$, resulting in strong spin relaxation at low fields and suppressed spin relaxation at high fields. Due to the presence of $\overline{B}_{ex,y}$   in equation \ref{eq:relax}, ferromagnetic ordering will produce a dip in  $R_{NL}$ that is centered away from zero and is hysteretic, while paramagnetic ordering will produce a non-hysteretic dip centered at zero field. Thus, the magnetic moments measured in these experiments are paramagnetic. The total longitudinal spin lifetime, $T^{total}_1$, of conduction electrons is dependent on both the usual spin relaxation due to spin orbit coupling ($\tau_{so}$) and longitudinal spin relaxation from the exchange field ($\tau^{ex}_1$), such 
that $\left( T^{total}_1 \right)^{-1}= \left( \tau^{ex}_1 \right)^{-1} +\left( \tau^{so} \right)^{-1}$. We apply the above model to the non-local spin transport data presented in Fig.~\ref{fig:main}f (sample A) and fit using equation (\ref{eq:nonlocal}), $\lambda = \sqrt{D T^{total}_1}$, and equation (\ref{eq:relax}) \cite{supplement}. The resulting fit (red line in Fig.~\ref{fig:analysis}b) replicates the shape and magnitude of the dip measured in $R_{NL}$ (black line in Fig.~\ref{fig:analysis}b). The field dependent  $T_1^{total}$ (Fig.~\ref{fig:analysis}c), exhibits a minimum of 464 ps at zero field and increases asymptotically towards $\tau_{so}$ = 531 ps for large $B_{app,y}$. The values obtained for $\Delta B$ and $\tau_c$ are 6.78 mT and 192 ps, respectively. The field-dependent spin relaxation following atomic hydrogen exposure, which emerges as a dip in $R_{NL}$, is a clear signature of paramagnetic moment formation.

Spin precession measurements provide further evidence for the presence of magnetic moments. Figure \ref{fig:analysis}d shows spin precession data for sample A (8 s exposure, $V_G-V_D$ = -15 V) with FM electrodes in the parallel alignment state. The Hanle curve has considerably narrowed compared to the precession measurements obtained prior to hydrogen adsorption (Fig.~\ref{fig:main}d). The sharpening of the Hanle curve results from the presence of an exchange field. The injected spins precess around a total field $B_{tot}=B_{app,z}+\overline{B}_{ex,z}$  (along $z$-axis) that includes not only the applied field, but also the exchange field from the paramagnetic moments. At 15 K and $B_{app,z} <$ 100 mT, the magnetization is proportional to the applied field so that 
$\overline{B}_{ex,z}=kB_{app,z}$, where $k$ is a proportionality constant. Thus, the spins precess about $B_{tot}$ with frequency $\omega = g_e \mu_B B_{tot}/\hbar = g_e (1+k) \mu_B B_{app,z}/\hbar =   g_e^* \mu_B B_{app,z}/\hbar$. To properly account for the enhanced $g$-factor induced by the magnetic moments, the Hanle equation must be modified to 
\begin{equation}
R_{NL}=S \int_{0}^{\infty} \frac{e^{-L^2/4Dt}}{\sqrt{4 \pi D t}}
\cos \left( \frac{g_e^* \mu_B B_{app,z}t}{\hbar}  \right)
e^{-t/T_2^{total}} dt
\label{eq:hanle}
\end{equation}	
where $T_2^{total}$  is the transverse spin lifetime. As shown in Fig.~\ref{fig:analysis}c, the $T_2^{total}$ is related to, but different from $T_1^{total}$ \cite{supplement}. Using the field dependent $T_2^{total}$, the precession data (red circles of Fig.~\ref{fig:analysis}d) is fit to equation \ref{eq:hanle} (black line) to yield a value of $g_e^*=7.13$. Physically, $g_e^*>2$ corresponds to an enhanced spin precession frequency resulting from the exchange field. A detailed discussion of the Hanle fitting and the gate-dependent properties of the exchange field are provided in the supplementary information 
\cite{supplement}. The dramatic narrowing of the Hanle peak combined with the emergence of a dip in $R_{NL}$ provides the most direct evidence to date for the formation of magnetic moments in graphene due to the adsorption of atomic hydrogen.

We now turn our attention to lattice vacancy defects in graphene. Several theoretical works suggest the similarity of magnetism due to vacancies and hydrogen-doping 
\cite{yazyev:2007,soriano:2011}, as both should create magnetic moments inside graphene due to the removal/hybridization of $p_z$ orbitals. To produce lattice vacancies in pristine SLG spin valves, we perform Ar-sputtering at low energies and examine the subsequent non-local spin transport. We again observe the emergence of a dip in $R_{NL}$ and narrowed Hanle curve, indicating the formation of paramagnetic moments in graphene \cite{supplement}. Given the very different chemical and structural properties of lattice vacancies compared to adsorbed hydrogen, the observation of similar features in the spin transport data provide strong evidence that the magnetic moments are created by the removal of $p_z$ orbitals from the $\pi$-band, as predicted theoretically.

In conclusion, clear signatures of magnetic moment formation are observed in both the non-local spin transport and Hanle precession data, which emerge only after exposure to atomic hydrogen or lattice vacancies. The results and techniques presented here are important for future developments in magnetism and spintronics.

We acknowledge Z. Zhao and C. N. Lau for their technical assistance and support from NSF (DMR-1007057, MRSEC DMR-0820414), ONR (N00014-12-1-0469), NRI-NSF (NEB-1124601) and DFG (SFB 689).

\bibliography{hydrogen_prl_response.bib}

\end{document}



\title{Supplemental material for: \\
Magnetic Moment Formation in Graphene Detected by Scattering of Pure Spin Currents}

\author{Kathleen M. McCreary}
\author{Adrian G. Swartz}
\author{\\Wei Han}
\author{Jaroslav Fabian}
\author{Roland K. Kawakami}

\date{\today}



\maketitle

\tableofcontents

\section{Atomic hydrogen exposure}  
\label{sec1}

A commercial Omicron source is used to expose SLG spin valve devices to atomic hydrogen at 15 K. Diatomic hydrogen is cracked inside a tungsten capillary tube that is heated by electron bombardment. The amount of hydrogen introduced to the chamber is controlled via a leak valve, which is tuned to maintain a chamber pressure of $1\times10^{-6}$ torr (the base pressure of the chamber is below $1\times10^{-9}$ torr). The heating power of the Omicron source is determined by the high voltage (HV) applied to the capillary and the emission current between capillary and filament (I$_{\text{em}}$). We use the parameters HV=1 kV and I$_{\text{em}}$=80 mA. The distance from source to sample is 100 mm. A shutter positioned between the SLG spin valve and hydrogen source is used in order to control the exposure time. Additionally, deflector plates are used to steer any charged ions away from the sample. 

As discussed in the main text, the exposure of SLG spin valves to atomic hydrogen substantially modifies charge transport properties, such as $\rho_G$ and $\mu$ (Fig. 1 of main text). We obtain an order of magnitude estimate for the hydrogen concentration based upon the changes in charge transport properties assuming adsorbed hydrogen induces resonant scattering. Comparing with previous experimental work on resonant scattering in graphene via fluorine doping \cite{Hong:2011} and lattice vacancies \cite{Chen:2009}, the hydrogen concentration is estimated to be on the order of 0.1\% for 8 s hydrogen exposure to sample A. This indicates samples are in the dilute limit of hydrogen coverage. 
 
 
\section{Control experiments to determine the origin of the dip in $R_{NL}$}  
\label{sec2}

\subsection{ Excluding field-dependent resistivity effects}   
\label{subsec2A}
Equation (1) of the main text shows that changes in the resistivity of graphene could change the spin density at the detector electrode. Thus, magnetic field dependent changes to the graphene resistivity could in principle produce variations in $R_{NL}$. To investigate this possibility in hydrogen-doped graphene, gate dependent $\rho_G$ measurements are performed on sample B at three distinct in-plane magnetic fields (Fig. \ref{fig:S1}). The three curves are indistinguishable, showing that the applied magnetic field has no effect on the measured $\rho_G$ of hydrogen-doped samples. Therefore, the dip in $R_{NL}$ does not originate from field-dependent changes in resistivity. 

\subsection{Excluding nuclear spin effects}  
\label{subsec2B}
We investigate the possibility that the observed dip in the non-local resistance originates from hyperfine coupling between conduction electrons and nuclear spins. In graphene, this scenario is unlikely due to the small abundance of intrinsic nuclear spins in carbon ($>$98\% of carbon is $^{12}$C, which has no nuclear spin) and a lack of contact hyperfine coupling in the $p_z$-orbitals that make up the conduction and valence bands. Nevertheless, investigating the situation is necessary because the adsorption of hydrogen on graphene may alter the hyperfine coupling. Two effects that could in principle alter the non-local resistance include hyperfine coupling to dynamically polarized nuclear spins \cite{Chan:2009,Salis:2009}  and organic magnetoresistance (OMAR) \cite{Nguyen:2007,Nguyen:2010}.

\begin{figure}
\includegraphics[scale=1.1]{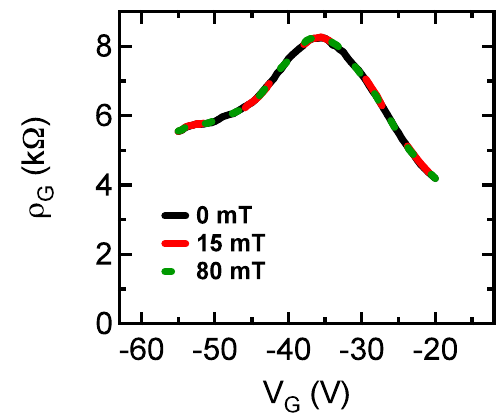}
\caption{\label{fig:S1} Investigating the source of the dip measured in $R_{NL}$. Gate-dependent $\rho_G$ of SLG after hydrogen exposure measured at several discrete in-plane applied magnetic fields. The applied field has no effect on the charge transport behavior of the graphene device.}
\end{figure}
\begin{figure}
\includegraphics[scale=1]{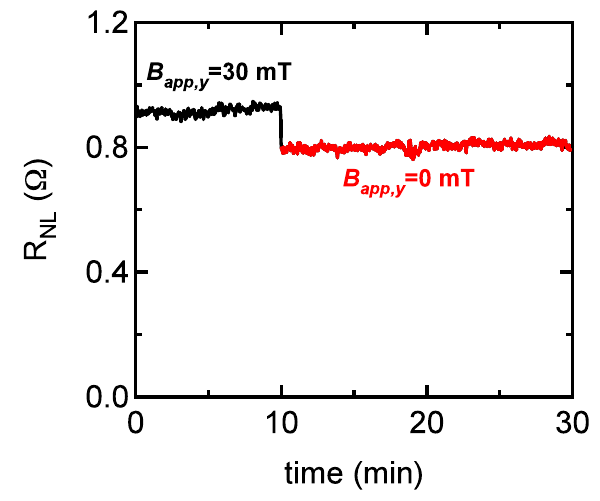}
\caption{\label{fig:S2} Test of nuclear spin effects. The non-local spin transport signal is measured for 10 min while $B_{app,y}$=30 mT (black). This ensures any possible nuclear spins will reach steady state. The field is ramped to zero, and held constant for the following 20 min (red). The observed response excludes hyperfine coupling to nuclear spins as the origin of the dip in $R_{NL}$.  }
\end{figure}

The effect of OMAR has previously been observed in carbon C$_{60}$ and functionalized carbon-based polymers \cite{Nguyen:2007,Nguyen:2010}. OMAR originates from hyperfine induced spin mixing between singlet and triplet states and manifests itself as a magnetic field dependent resistivity. However, as shown in Fig. \ref{fig:S1}, resistivity does not change as a function of applied in-plane magnetic field. This confirms OMAR is not responsible for the observed dip in $R_{NL}$. 

The effect of dynamic nuclear polarization (DNP) was demonstrated clearly by Salis {\it{et al.}} \cite{Salis:2009} and Chan {\it{et al.}} \cite{Chan:2009} who investigated GaAs non-local spin valves at low temperatures. Specifically, Salis et al. observed a dip in $R_{NL}$ at zero applied magnetic field, similar to the dip we observe in hydrogen-doped graphene. They attributed their dip to precessional spin dephasing caused by hyperfine coupling to dynamically polarized nuclear spins. To determine whether such nuclear spin effects are present in hydrogen-doped graphene we perform a series of tests. 

{\underline{{\bf{Test 1:}}}} Nuclear spin relaxation times are typically long ($\sim$minutes), and therefore, slow dynamics at this time scale are a characteristic of effects related to DNP. This is manifested in non-local spin transport data as a ``lab time'' dependence \cite{Salis:2009}. In our investigation of hydrogen-doped graphene, we do not observe a lab time dependence or a magnetic field ramp rate dependence.  

{\underline{{\bf{Test 2:}}}} A characteristic feature of hyperfine coupling to the nuclear spin bath through DNP is the nuclear field's linear dependence on the applied field \cite{Salis:2009}. Specifically, at zero field the nuclear spin bath depolarizes slowly over time. The depolarization of the nuclear spin bath is evident in $R_{NL}$ data as the gradual decrease and eventual disappearance (after a few minutes) of the dip when the applied field is set to zero (Figure 1 of Salis {\it{et al.}} \cite{Salis:2009}). We perform this test on hydrogen-doped graphene spin valves, as shown in Fig. \ref{fig:S2}. First, the magnetic field is held at -30 mT for 10 minutes to ensure any possible nuclear spin transients reach steady state. The $R_{NL}$ is measured continuously during this period and exhibits a value of $\sim$0.92 $\Omega$. Then the magnetic field is quickly reduced to zero, coinciding with the immediate drop of $R_{NL}$ to $\sim$0.80  $\Omega$. This drop occurs because $B_{app,y}$=0 mT is at the center of the dip in $R_{NL}$. The $R_{NL}$ is measured over the following 20 minutes, for which the observed value remains unchanged at $\sim$0.80  $\Omega$. This indicates that the magnitude of the dip is independent of lab time. If the dip were due to hyperfine coupling to dynamically polarized nuclear spins, then the magnitude of the dip would gradually decrease to zero (i.e. $R_{NL}$ would increase back to $\sim$0.92  $\Omega$). Because this behavior is not observed, the dip in $R_{NL}$ cannot be due to DNP.  

{\underline{{\bf{Test 3:}}}} For the case of DNP, applying a constant out-of-plane magnetic field during a non-local spin transport measurement (in-plane field scan) results in a characteristic feature of nuclear depolarization and repolarization as the in-plane field crosses zero (Figure 1 of Chan {\it{et al.}} \cite{Chan:2009}). We perform this measurement on hydrogen-doped graphene and observe no evidence of depolarization/repolarization features. Together, these three tests show that the dip in $R_{NL}$ is not due to hyperfine coupling to dynamically polarized nuclear spins. 

The above investigations of OMAR and DNP conclusively exclude the possibility of hyperfine coupling with nuclear spins as the source of the observed dip in the non-local spin transport. 

\subsection{Excluding changes in magnetization of the electrodes} 
\label{subsec2C}

\begin{figure*}[t]
\includegraphics[scale=.9]{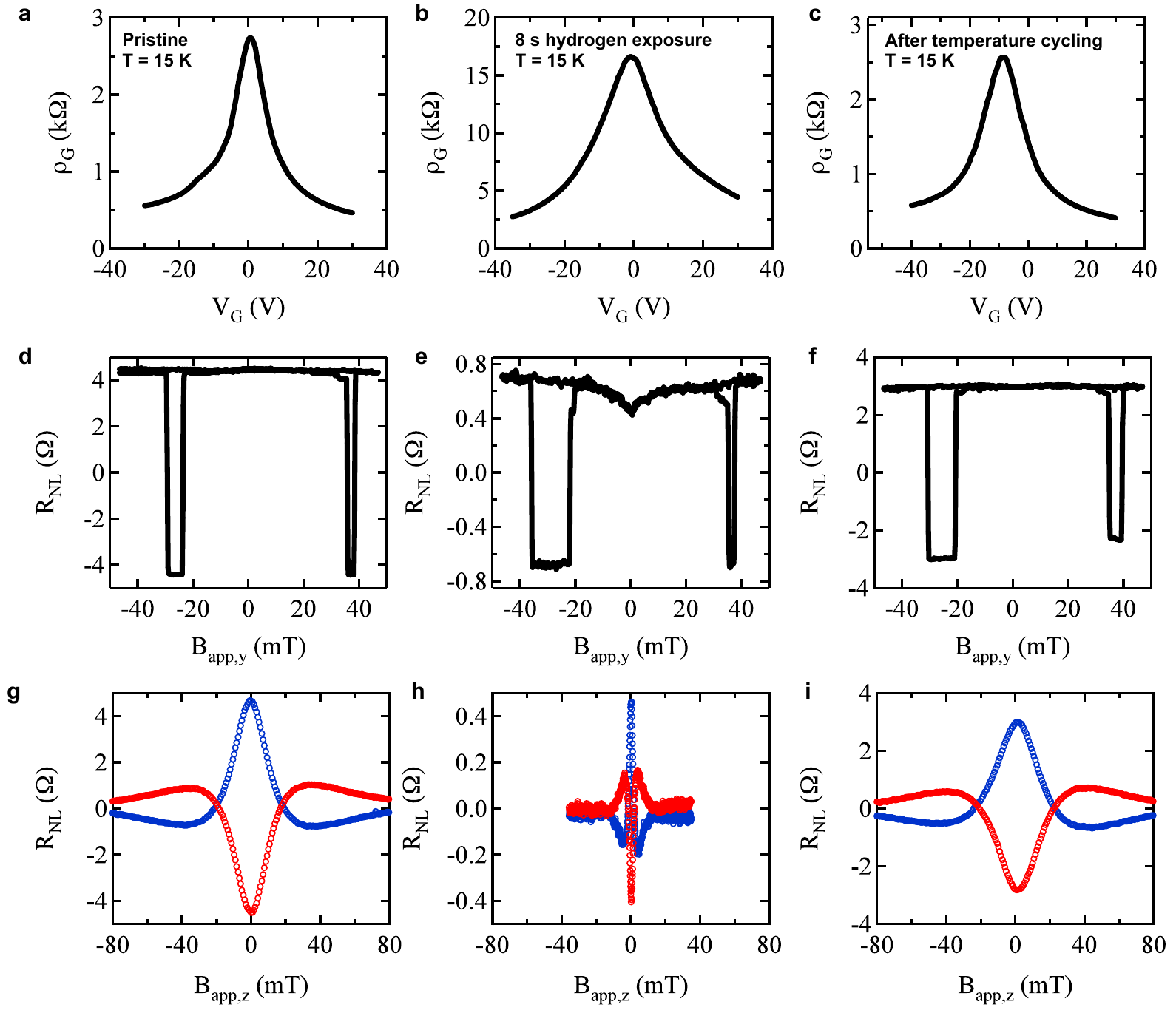}
\caption{\label{fig:S3} Reversibility of effects due to hydrogen doping of SLG. a, Gate dependent resistivity for pristine SLG at T=15 K. Data set corresponds to sample A. b, Gate dependent resistivity after 8 s hydrogen exposure at T=15 K. c, Gate dependent resistivity at T=15 K, measured after hydrogen exposure at 15 K followed by warming to room temperature and subsequent cooling to 15 K. d, $R_{NL}$ for pristine SLG. e, $R_{NL}$ after 8 s hydrogen exposure. f, $R_{NL}$ after temperature cycling. g, Hanle spin precession data for pristine SLG. Blue (red) data is for parallel (antiparallel) magnetizations. h, Hanle spin precession data after 8 s hydrogen exposure. i, Hanle spin precession data after temperature cycling. All $R_{NL}$ and spin precession data are measured at $V_G-V_D$=-15 V.}
\end{figure*}

Here we examine the possibility that the observed behavior in $R_{NL}$ after hydrogen exposure is caused by spurious changes to the magnetic electrodes.  It is known that hydrogen adsorption onto ferromagnetic materials can alter their magnetic properties \cite{Dietz:1959,Boudart:1977,Vollmer:1999,Weinert:1985,Nordlander:1984}. In the present experiment, following the standard fabrication procedure for tunneling contacts to graphene as discussed in previous work \cite{Vollmer:1999}, the magnetic electrodes are capped with 5 nm Al$_2$O$_3$. Also, we note that in the $R_{NL}$ data presented in this work, the coercive fields of the electrodes remain unchanged after hydrogen exposure.  

Further it is known that the bonding energy for hydrogen adsorption on transition metals is several eV ($\sim$2.6 eV for Co \cite{Nordlander:1984}), suggesting that H-Co chemisorption would be robust to temperature cycling. Here we note that the observed behavior in $R_{NL}$ and spin precession data after the introduction of atomic hydrogen at T=15 K are reversible upon temperature cycling. At cryogenic temperatures, exposure of SLG graphene spin valves to hydrogen dramatically alters the charge and spin properties. The gate dependent resistivity increases, whereas the magnitude of $R_{NL}$ decreases and exhibits a dip at low field. The observed dip is due to an increase in the spin scattering. Also, precession measurements exhibit a narrowing of the Hanle curve. Next, the sample is warmed to room temperature and subsequently re-cooled to T=15 K. The spin and charge transport properties are then re-measured. We have found that the charge transport properties of the gate dependent resistivity and mobility nearly recover to the pristine values. For instance, for sample A, the mobility is $\mu$=6105 cm$^2$/Vs for pristine graphene.  After 8 s hydrogen exposure the mobility decreases to $\mu$=495 cm$^2$/Vs. After temperature cycling, the mobility recovers to $\mu$=5450 cm$^2$/Vs. The recovery indicates the effect of hydrogen is removed through desorption, cluster formation, or a combination of the two. After temperature cycling, the spin signal $R_{NL}$ contains no indication of a dip near low field and has increased to 5.95 $\Omega$. Also, the spin precession data broadens and conventional Hanle analysis (equation \ref{S14}) yields $\tau^{so}$=353 ps and $D$=0.022 m$^2$/s. As mentioned above, changes in the Co electrodes should persist after temperature cycling and so cannot explain the observed behavior. Upon re-hydrogenation at cryogenic temperatures the key features (dip in $R_{NL}$, narrowed Hanle curve) return. \mbox{Fig. \ref{fig:S3}} summarizes the changes in resistivity, $R_{NL}$, and spin precession upon temperature cycling. 

\begin{figure}
\includegraphics[scale=1]{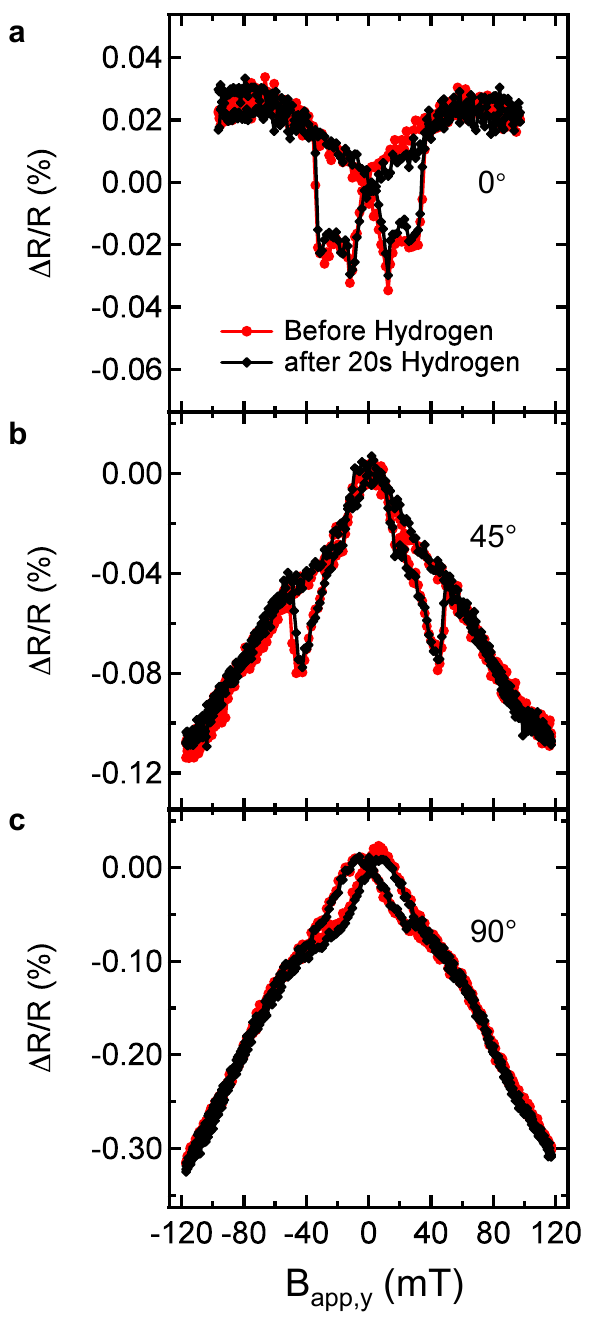}
\caption{\label{fig:S4} Magnetoresistance of Co wires before and after hydrogen exposure at T=10 K. a, $\Delta$R/R (\%) of a single Co wire measured before hydrogen exposure (red circles) for a wire oriented along $B_{app,y}$ (0 $^\circ$). The wires were fabricated using standard e-beam lithography and angle evaporation in the same fashion as for producing tunnel barrier electrodes on SLG. The observed features are due to anisotropic magnetoresistance (AMR) and reflect the magnetization of the Co wire. No effect is observed after 20 s of hydrogen exposure (black diamonds). b, $\Delta$R/R (\%) of a Co wire oriented 45 $^\circ$ with respect to $B_{app,y}$ before hydrogen exposure (red circles) and after 20 s exposure (black diamonds). c, $\Delta$R/R (\%) of a Co wire oriented 90 $^\circ$ with respect to $B_{app,y}$ before hydrogen exposure (red circles) and after 20 s exposure (black diamonds).  In all cases the wires are oriented in-plane. 
}
\end{figure}

To comprehensively verify that changes in magnetism of the Co electrodes is not the cause of the observed dip, we fabricate Co wires in the same fashion as for tunnel barrier contacts to graphene. Three 300 nm wide Co wires with equal length ($L$=200 $\mu$m) and 80 nm thick are oriented at 0 $^\circ$, 45 $^\circ$, and 90 $^\circ$ with respect to an in-plane applied field. Figure \ref{fig:S4} shows the magnetoresistance plotted as $\Delta$R/R (\%) vs. $B_{app,y}$ for the three different orientations before and after 20 s hydrogen exposure.  We note that 20 s hydrogen exposure is significantly more than typically needed to generate observable effects in $R_{NL}$. As can be seen in Fig. \ref{fig:S4}, no effect is seen in the magnetoresistance of the Co wires upon hydrogen exposure. The switching fields and shape remain unchanged along with the anisotropy behavior. Therefore, we rule out effects of alterations in the Co wire magnetization as a possible source for the origin of the observed dip at zero field in $R_{NL}$. 
 
 
\section{Effective exchange field model}  
\label{sec3}
To quantitatively model the experiment, we consider electron spins $\vec{S}_e$ moving in an effective magnetic field of randomly positioned local magnetic moments $\vec{S}_M$ of filling density $\eta_M$. Each electron feels the average spin interaction  

\begin{eqnarray}
\nonumber
H_e&&=\eta_M A_{ex} \vec{S}_e \cdot \langle \vec{S}_M \rangle + g_e \mu_B \vec{S}_e \cdot \vec{B}_{app} 
\\ \label{S1}
&&=g_e \mu_B \vec{S}_e \cdot (\overline{\vec{B}}_{ex} + \vec{B}_{app})
\end{eqnarray}
and
\begin{equation}
\label{S2}
\overline{\vec{B}}_{ex} = \eta_M A_{ex} \langle \vec{S}_M \rangle / g_e \mu_B
\end{equation}
where $\overline{\vec{B}}_{ex}$ is the effective exchange magnetic field, $g_e$ is the electron g-factor, $\mu_B$ is the Bohr magneton, and $A_{ex}$  is the strength of the exchange coupling between $\vec{S}_e$ and $\vec{S}_M$. The averaging $\langle...\rangle$ is over the ensemble of magnetic moments. The effective exchange field $\overline{\vec{B}}_{ex}$ contributes to the Larmor frequency and enhances the electron g-factor.

As the spins diffuse through the lattice they experience varying magnetic moments which results in varying Larmor frequencies. In the local frame associated with the electrons this can be described by a time-dependent, randomly fluctuating magnetic field, 
\mbox{$\vec{B}_{ex}(t)\!=\!\overline{\vec{B}}_{ex}\!+\!\Delta \overline{\vec{B}}_{ex}(t)$}
with the rms value given by the time average
\begin{eqnarray}
\label{S3}
&&(\Delta B^{rms}_{ex})^2_\alpha \!=\! \langle [\Delta B_{ex,\alpha}(t)]^2 \rangle_t
\\ \nonumber
\\ \label{S4}
&&(\Delta B^{rms}_{ex})^2 \!=\! (\Delta B^{rms}_{ex})^2_x + (\Delta B^{rms}_{ex})^2_y + (\Delta B^{rms}_{ex})^2_z
\end{eqnarray}
where $\alpha$ is an $xyz$ component index. The time scale of the fluctuation is given by a correlation time $\tau_c$ defined by   
\begin{equation}
\label{S5}
\langle \Delta \vec{B}_{ex}(t) \cdot \Delta \vec{B}_{ex}(t\!-\!t') \rangle_t \propto \text{exp}(-t'/\tau_c)
\end{equation}

Spin relaxation resulting from a randomly fluctuating magnetic field has been solved in the review article by \citet[Section IV.B.2]{Fabian:2007} and is mathematically analogous to the D'yakanov-Perel model\cite{DP:1971}. For the non-local spin signal geometry, the injected spin polarization and the applied magnetic field lie along the same axis ($y$-axis) and the spin relaxation rate is given by longitudinal spin relaxation equation IV.36 of \citet{Fabian:2007} The equation is rewritten using
$\omega_\alpha\!=\!g_e \mu_B \Delta B_{ex,\alpha}/\hbar$
and
$\omega_0\!=\!g_e \mu_B \overline{B}_{total} / \hbar\!=\!g_e \mu_B(B_{app,y} + \overline{B}_{ex,y}/\hbar)$
to yield,
\begin{equation}
\label{S6}
\frac{1}{\tau_1^{ex}} = \frac{(\Delta B)^2}{\tau_c}\frac{1}{(B_{app,y}+\overline{B}_{ex,y})^2+\left(\frac{\hbar}{g_e \mu_B \tau_c}\right)^2}
\end{equation}
where $(\Delta B)^2 \!=\! (\Delta B_{ex})^2_z \!+\! (\Delta B_{ex})^2_z$. In other words, precession around randomly fluctuating exchange fields along the $x$- and $z$-axes induce spin relaxation. Equation (\ref{S6}) also shows that the spin relaxation is suppressed by a large applied magnetic field. Intuitively, this occurs because the precession axis is defined by the large applied field (along $y$-axis) and fluctuating fields along the $x$- and $z$-axes have very little ability to tilt the precession axis. This peak in spin relaxation at low magnetic fields produces the observed dip in $R_{NL}$. 

The presence of the average exchange field $\overline{B}_{ex,y}$ in equation (\ref{S6}) shows that the spin relaxation is maximized when the $B_{app,y} \!+\! \overline{B}_{ex,y} \!=\! 0\text{, or } B_{app,y} \!=\! -\overline{B}_{ex,y}$. Because $\overline{B}_{ex,y}$, is proportional to the magnetization, paramagnetic moments will generate a dip in $R_{NL}$ centered at zero applied field while ferromagnetic ordering will generate a hysteretic dip centered away from zero applied field. The observed dip (Fig. 1e and 1f of main text) is centered at zero applied field and is not hysteretic, signifying the observed magnetic moments are paramagnetic. 

For paramagnetic moments, $\overline{B}_{ex,y}$ takes the form of the Brillouin function ($B_J$) and is given by 
\begin{eqnarray}
\nonumber
\overline{B}_{ex,y}=&&\ \eta_M A_{ex} \langle S_{M,y} \rangle / g_e \mu_B\\
\label{S7}
=&&\ \eta_M A_{ex} J B_J(\xi) / g_e \mu_B\\  \nonumber
\\
\label{S8}
B_J(\xi) =&& \frac{2J + 1}{2J}\text{coth}\!\left(\frac{2J + 1}{2J}\xi\!\right)-\frac{1}{2J}\text{coth}\!\left(\frac{1}{2J}\xi\!\right)
\end{eqnarray}
where $J$ is the total angular momentum quantum number of the magnetic moment, \mbox{$\xi\!=\!\frac{J g_e \mu_B}{k_B \text{T}}B_{app,y}$}, $k_B$ is BoltzmannÕs constant, and T is temperature. For our experiments at T=15 K, this reduces to \mbox{$\xi\!=\!J g_e B_{app,y}/(22.32\text{ Tesla})$}. Thus, for the values of $B_{app,y}$ in our experiments $\xi \ll 1$ so that $B_J$$\approx$$(J+1)\xi/3J $ to yield 
\begin{equation}
\label{S9}
\overline{B}_{ex,y} = \frac{\eta_M A_{ex} J(J+1)}{3\mu_B}\left(\frac{B_{app,y}}{22.32\text{ Tesla}}\right)
\end{equation}
Thus, the total field can be written as
\begin{eqnarray}
\nonumber
&&B_{total} = B_{app,y} + \overline{B}_{ex,y}\\ \nonumber \\ \nonumber
&&=B_{app,y} + \frac{\eta_M A_{ex} J(J+1)}{3\mu_B}\!\!\left(\frac{B_{app,y}}{22.32\text{ Tesla}}\right) \\ \nonumber \\
\label{S10}
&&=\left(\!1+\frac{\eta_M A_{ex} J(J+1)}{3\mu_B(22.32\text{ Tesla})}\!\right)\!\!B_{app,y} = \frac{g_e^*}{g_e} B_{app,y}
\end{eqnarray}
where the $g_e^*$ is the enhanced g-factor due to the presence of the exchange field. Substituting this into equation (\ref{S6}) yields the expression for spin relaxation from para\-magnetic moments in the linear regime,   
\begin{eqnarray}
\nonumber
\frac{1}{\tau_1^{ex}} &&= \frac{(\Delta B)^2}{\tau_c}\frac{1}{\left(\frac{g_e^*}{g_e} B_{app,y}\right)^2+\left(\frac{\hbar}{g_e \mu_B \tau_c}\right)^2}
\\ \nonumber
\\ \label{S11}
&&= \frac{\frac{(\Delta B)^2}{\tau_c}\!\left(\frac{g_e}{g_e^*}\right)^2}{(B_{app,y})^2+\left(\frac{\hbar}{g_e^* \mu_B \tau_c}\right)^2}
\end{eqnarray}
Thus, the longitudinal spin relaxation rate is a Lorentzian with a peak at zero applied field. 

For the Hanle geometry, the injected spin polarization is along the $y$-axis and the applied magnetic field is along the $z$-axis. In this case the spin relaxation rate is given by transverse spin relaxation equation IV.38 in \citet{Fabian:2007} The equation is rewritten using
\mbox{$\omega_\alpha = g_e \mu_B \Delta B_{ex,\alpha}/\hbar$} (for \mbox{$\alpha\!=\!x,y,z$}) and \mbox{$\omega_0 = g_e \mu_B \overline{B}_{total}/\hbar$} = $g_e \mu_B (B_{app,z} + \overline{B}_{ex,z})/\hbar$
to yield
\begin{eqnarray}
\nonumber
\frac{1}{\tau_2^{ex}} &&= \frac{1}{2}\!\!\left[\frac{(\Delta B)^2}{\tau_c\left(\frac{\hbar}{g_e \mu_B \tau_c}\right)^2}\right]
\\ \nonumber
\\ \label{S12}
&&+ \frac{1}{2}\!\!\left[\frac{(\Delta B)^2}{\tau_c}\frac{1}{(B_{app,z}+\overline{B}_{ex,z})^2+\left(\frac{\hbar}{g_e \mu_B \tau_c}\right)^2}\right]
\end{eqnarray} 
where the fluctuating field is assumed to be isotropic: $(\Delta B_{ex})^2_x = (\Delta B_{ex})^2_y = (\Delta B_{ex})^2_z $. For para\-magnetic moments, this becomes 
\begin{eqnarray}
\label{S13}
\frac{1}{\tau_2^{ex}}=\!\frac{1}{2}\!\!\!\left[\!\frac{\frac{(\Delta B)^2}{\tau_c}\!\left(\!\frac{g_e}{g_e^*}\!\right)^2}{\left(\!\frac{\hbar}{g_e^* \mu_B \tau_c}\!\right)^2}\!\right]\!+\!\frac{1}{2}\!\!\!\left[\!\frac{\frac{(\Delta B)^2}{\tau_c}\!\left(\!\frac{g_e}{g_e^*}\!\right)^2}{(B_{app,z})^2\!+\!\left(\!\frac{\hbar}{g_e^* \mu_B \tau_c}\!\right)^2}\!\right]
\end{eqnarray} 
%


\section{Application of the exchange field model to experimental {\it{R}}$_{NL}$ and Hanle data}  
\label{sec4}
The spin relaxation, diffusion coefficient, and interfacial spin polarization of the pristine sample are determined though analysis of $R_{NL}$ and spin precession measurements. Fitting of spin precession data to the Hanle equation 
\begin{equation}
\label{S14}
R_{NL}\!=\!S\!\int \limits_{0}^{\infty}\!\frac{e^{-L^2/4Dt}}{\sqrt{4 \pi D t}}\!\cos\!\left(\!\frac{g_e \mu_B}{\hbar}B_{app,z}t\!\right)\!e^{-t/\tau^{so}} dt
\end{equation}	
provides values of spin lifetime ($\tau^{so}$), diffusion coefficient ($D$), Hanle amplitude ($S$), and spin diffusion length 
\mbox{$(\lambda\!=\!\sqrt{D\,\tau^{so}})$}.
For pristine graphene, the electron g-factor, $g_e$, is assumed to be 2. It should be noted that in the case of pristine graphene, the longitudinal spin relaxation ($\tau_1^{so}$) and transverse spin relaxation ($\tau_2^{so}$) due to spin orbit coupling are equivalent 
\mbox{$(\tau_1^{so}\!=\!\tau_2^{so}\!=\!\tau^{so})$} \cite{Tombros:2007}. 
Data measured on sample A (Fig. 1d of main text), yields $\tau^{so}$=479 ps, $D$=0.023 m$^2$/s and $\lambda$=3.3 $\mu$m for the channel length, $L$=5.25 $\mu$m. The corresponding $R_{NL}$ data is fit with the non-local resistance equation (\ref{S15}) to obtain the interfacial spin polarization, $P_J$, for the graphene device. 
\begin{eqnarray}
\nonumber
R^{(P/AP)}_{NL}\!=\pm 2R_Ge^{-L/\lambda} \prod^{2}_{i=1}\!
\left(\!\frac{P_J \frac{R_i}{R_G}}{1-P_J^2}\!+\!\frac{P_F \frac{R_F}{R_G}}{1-P_F^2}\!\right)
\\ \label{S15}
\times
\left[\prod^{2}_{i=1}\!
\left(\!1\!+\!\frac{2 \frac{R_i}{R_G}}{1-P_J^2}\!+\!\frac{2 \frac{R_F}{R_G}}{1-P_F^2}\!\!\right)\!-\!e^{-2L/\lambda}\right]^{-1}
\end{eqnarray}

In the above equation, $R_G$ is the graphene spin resistance defined by, \mbox{$R_G\!=\!\rho_G\lambda/w$}, where $\rho_G$ is the resistivity and $w$ is the graphene width.  $R_{1,2}$ denotes the contact resistances of injector and detector electrodes, $P_F$ is the ferromagnetic electrode spin polarization (assumed to be 0.35 for cobalt), and 
\mbox{$R_F\!=\!\rho_F\frac{\lambda_F}{l_jw}$}
is the spin resistance of the ferromagnet, where $\rho_F\!=\!5.8\times10^{-8}$ $\Omega$m is the resistivity of cobalt, $l_j$=50 nm is the effective spin injector contact length of the ferromagnetic electrode and is determined by the fabrication procedures (see \citet{Han:2012} for details), and lastly, $\lambda_F$ is the spin diffusion length of the ferromagnet, taken to be 38 nm in cobalt. 
The measured $\Delta R_{NL}$=8.8 $\Omega$ for Sample A (Fig. 1c) corresponds to a $P_J$=0.20 for parameters $\lambda$=3.3 $\mu$m, $\rho_G$=898 $\Omega$, $w$=2.3 $\mu$m, $R_1$=15.76 k$\Omega$, and $R_2$=4.00 k$\Omega$. The contact resistances are measured in a three terminal geometry \cite{Han:2010} and are found to be unaffected by hydrogen exposure. The measured value of 20\% interfacial spin polarization is comparable to previously reported values for efficient spin injection into SLG through tunneling contacts \cite{Han:2010}. $P_J$ is assumed to remain constant throughout hydrogen exposure, a reasonable assumption since the graphene at the site of spin injection is protected by the electrode, the hydrogen does not alter the 
cobalt (see section \ref{subsec2C}), and contact resistances remain unchanged.      

As discussed in the main text, exposure to atomic hydrogen results in the formation of magnetic moments, detected as a dip in $R_{NL}$. Additionally, a sharpened Hanle curve signifies enhanced precession of injected spins due to the presence of an exchange field caused by the moments. The exchange field is not accounted for in the standard Hanle equation (\ref{S14}), preventing direct determination of $\tau^{so}$ and $D$ through Hanle fitting. Instead, the Einstein relation 
\begin{equation}
\label{S16}
D=\frac{\sigma}{e^2\nu}
\end{equation}
is employed to obtain $D$ for hydrogen-doped samples, where $\nu$ denotes the density of states and $e$ is the electron charge. Assuming $\nu$ is unchanged by exposure to hydrogen, which is reasonable in the dilute limit, the diffusion coefficient of hydrogen-doped samples $D_{hyd}$ is determined from the pristine diffusion coefficient $D_{pristine}$ and the conductivities of hydrogen-doped ($\sigma_{hyd}$) and pristine ($\sigma_{pristine}$) graphene. 
\begin{equation}
\label{S17}
\frac{D_{hyd}}{D_{pristine}}=\left(\frac{\sigma_{hyd}}{e^2\nu}\right)\!/\!\left(\frac{\sigma_{pristine}}{e^2\nu}\right) = \frac{\sigma_{hyd}}{\sigma_{pristine}}
\end{equation}
The change in conductivity from 1.113 mS to 0.143 mS following hydrogen exposure results in $D_{hyd}$=0.0029 m$^2$/s. 

The longitudinal spin lifetime is evaluated by examining $R_{NL}$ for the hydrogen-doped sample. As shown in equation (\ref{S11}) the spin relaxation rate arising from the presence of magnetic moments is described by a Lorentzian centered at $B_{app,y}$=0, and can be fit using the general form
\begin{equation}
\label{S18}
\frac{1}{\tau_1^{ex}} = \Gamma \frac{\gamma^2}{(B_{app,y})^2 + \gamma^2}
\end{equation}
The total longitudinal spin lifetime, $T_1^{total}$, depends on both $\tau_1^{ex}$ and $\tau^{so}$ through the relation 
\begin{equation}
\label{S19}
\frac{1}{T_1^{total}} = \frac{1}{\tau^{so}} +\frac{1}{\tau_1^{ex}} 
\end{equation}
subsequently affecting the spin diffusion length $\lambda$
\begin{eqnarray}
\nonumber
\lambda &&= \sqrt{D\,T_1^{total}} = \sqrt{D\left( \frac{1}{\tau^{so}}+\frac{1}{\tau_1^{ex}}\right)^{-1}}
\\ \nonumber
\\ \label{S20}
&&= \sqrt{D\left( \frac{1}{\tau^{so}}+\Gamma \frac{\gamma^2}{(B_{app,y})^2 + \gamma^2}\right)^{-1}}
\end{eqnarray}
The field dependent $\lambda$ directly translates to a field dependence in the non-local resistance causing the experimentally observed dip in $R_{NL}$ at zero applied field. Values for $\tau^{so}$, $\Gamma$, and $\gamma$ are determined by fitting the measured $R_{NL}$ data to the non-local resistance equation (\ref{S15}), where $\lambda$ is field dependent and defined by equation (\ref{S20}). The best fit to sample A is obtained by $\tau^{so}$=531 ps, \mbox{$\Gamma$=$2.73\!\times\!10^8$ s$^{-1}$}, and $\gamma$=8.32 mT and is displayed as the red curve in Fig. 3a (main text), using, $R_F$=0.019 $\Omega$, $\rho_G$=6.99 k$\Omega$ $R_1$=15.76 k$\Omega$, $R_2$=4.00 k$\Omega$, $P_J$=0.20, $P_F$=0.35, and $L$=5.25 $\mu$m. The field dependent values of $T_1^{total}$ are displayed as the red curve in Fig. 3b (main text). Clearly, this model explains the data well and may also be relevant for dip features observed recently in metallic lateral spin valves \cite{Mihajlovic:2011}. 

Following the determination of $\tau^{so}$, $\Gamma$, and $\gamma$, the spin precession data for the hydrogen-doped sample is analyzed in order to obtain values for $g_e^*$. The standard Hanle equation (\ref{S14}) must be modified to account for precession induced by both the applied field and the exchange field ($B_{ex,z}$) produced by magnetic moments as well as the field dependent transverse spin lifetime $T_2^{total}$, where \mbox{$\frac{1}{T_2^{total}}\!=\!\frac{1}{\tau^{so}}\!+\!\frac{1}{\tau_2^{ex}}$}. For the Hanle geometry, the spin relaxation rate from the magnetic moments is given by equation (\ref{S13}). Thus, in terms of the Lorentzian parameters $\Gamma$ and $\gamma$, the total spin relaxation rate is, 
\begin{equation}
\label{S21}
\frac{1}{T_2^{total}} = \frac{1}{\tau^{so}} + \frac{\Gamma}{2}\left(1+\frac{\gamma^2}{(B_{app,z})^2 + \gamma^2}\right)
\end{equation}
For comparison, Fig. 3b (main text) displays the total longitudinal spin lifetime $T_1^{total}$ (red curve) and total transverse spin lifetime $T_2^{total}$ (black curve) for hydrogen-doped graphene sample A.  

The modified Hanle equation is dependent on $T_2^{total}$ and takes the form 
\begin{eqnarray}
\nonumber
R_{NL}\!=\!S&&\int \limits_{0}^{\infty}\! \frac{e^{-L^2/4Dt}}{\sqrt{4 \pi D t}} 
e^{-t \left[
 \frac{1}{\tau^{so}} + \frac{\Gamma}{2}\left(1+\frac{\gamma^2}{(B_{app,z})^2 + \gamma^2}\right)\right]}
\\ \label{S22}
&&\times \cos\!\left( \frac{g_e \mu_B}{\hbar}(B_{app,z} + \overline{B}_{ex,z})t  \right) dt
\end{eqnarray}	
For paramagnetic moments, $\overline{B}_{ex,z}$ is described by the Brillouin function. Additionally, 
\mbox{$\frac{JB_{app,z}}{(22.32\text{Tesla})}\ll1$}, so that $\overline{B}_{ex,z}$ can be represented by the low field approximation 
\begin{equation}
\label{S23}
\overline{B}_{ex,z} = \frac{\eta_M A_{ex} J(J+1)}{3\mu_B}\left(\frac{B_{app,z}}{22.32\text{ Tesla}}\right)
\end{equation}
resulting in,
\begin{eqnarray}
\label{S24}
&&R_{NL}\!=\!S\!\int \limits_{0}^{\infty}\! \frac{e^{-L^2/4Dt}}{\sqrt{4 \pi D t}} 
e^{-t \left[ \frac{1}{\tau^{so}} + \frac{\Gamma}{2}\left(1+\frac{\gamma^2}{(B_{app,z})^2 + \gamma^2}\right)\right]}
\\ \nonumber
&&\times\cos\!\left[\!\frac{g_e \mu_B}{\hbar}\!\!
\left(\!\!B_{app,z}\!+\!\frac{\eta_M A_{ex}J(J\!+\!1)}{3\mu_B}\!\!\left(\!\!\frac{B_{app,z}}{22.32\text{ Tesla}}\!\right)\!\!\right)\!t \right]\! dt
\end{eqnarray}	
which simplifies to,
\begin{eqnarray}
\nonumber
R_{NL}\!=\!S&&\int \limits_{0}^{\infty}\! \frac{e^{-L^2/4Dt}}{\sqrt{4 \pi D t}} 
e^{-t \left[ \frac{1}{\tau^{so}} + \frac{\Gamma}{2}\left(1+\frac{\gamma^2}{(B_{app,z})^2 + \gamma^2}\right)\right]}
\\ \label{S25}
&&\times \cos\!\left( \frac{g_e^* \mu_B}{\hbar}B_{app,z}t  \right) dt
\end{eqnarray}	
with,
\begin{equation}
\label{S26}
g_e^* = g_e\!\left[1+\frac{\eta_M A_{ex} J(J+1)}{3 \mu_B (22.32\text{ Tesla})}    \right]
\end{equation}
Fitting of precession data to equation (\ref{S25}) yields values for Hanle amplitude ($S$) and $g_e^*$. In the fit, fixed parameters are: $\Gamma$=2.73$\times$10$^8$ s$^{-1}$ and $\gamma$=8.32 mT, as determined by analysis of non-local resistance, $D$=0.0029 m$^2$/s from the Einstein relation, and $L$=5.25 $\mu$m. The best fit of Hanle data (presented in Fig. 3c (main text)) results in a $g_e^*$ value of 7.13.  We take a reasonable value for the exchange coupling of $A_{ex}\sim1$ eV and the paramagnetic spin of $J=1/2$ expected for the unpaired electrons due to hydrogen adatoms on graphene.  Using the value $g_e^*$=7.13, we can independently estimate the fractional filling density of hydrogen induced magnetic moments to be $\eta_M\sim1$ \% using
equation (\ref{S26}). This is in reasonable agreement with the order of magnitude estimate of 0.1\% determined from the resistivity for 8 s hydrogen exposure to sample A.

The correlation time ($\tau_c$) and the rms fluctuations in exchange field ($\Delta B$) are determined by comparing equations (\ref{S11}) and (\ref{S18}) to give 
\begin{eqnarray}
\label{S27}
\tau_c &&= \frac{\hbar}{g_e^*\mu_B} \frac{1}{\gamma} 
\\ \nonumber
\\ \label{S28}
(\Delta B)^2 &&= \gamma \left(\frac{g_e^*}{g_e}\right) \frac{\hbar \Gamma}{g_e \mu_B}
\end{eqnarray}
Using $g_e^*$=7.13 from the Hanle fit and $\Gamma$=2.73$\times$10$^8$ s$^{-1}$  and $\gamma$=8.32 mT from the non-local fit, we obtain values of $\tau_c$=192 ps and \mbox{$(\Delta B)^2$=4.59$\times$10$^{-5}$ T$^2$} (or $\Delta B = \sqrt{(\Delta B_{ex,x})^2 + (\Delta B_{ex,z})^2}$=6.78 mT).


\section{Magnetic moments generated by lattice vacancies}  
\label{sec5}
\begin{figure}[b]
\includegraphics[scale=.85]{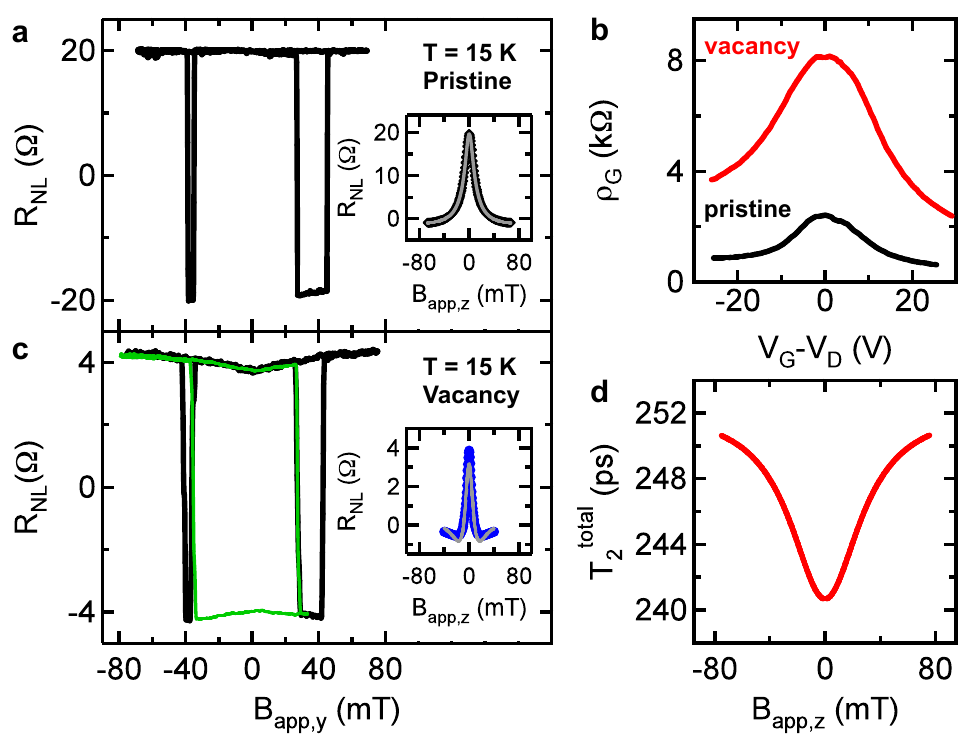}
\caption{\label{fig:S5}  The effect of lattice vacancies on charge and spin transport in SLG at 15 K. a, Non-local spin transport data in pristine graphene with $\Delta R_{NL}$=38.9 $\Omega$. Hanle precession data (inset) yields values of $\tau^{so}$= 859 ps and $D$=0.023 m$^2$/s. b, Gate dependent $\rho_G$ for pristine graphene (black) and following lattice vacancy formation via Ar-sputtering (red). c, Non-local spin transport data in SLG containing vacancies. After Ar-sputtering, $\Delta R_{NL}$ is decreased to 8.2 $\Omega$ and exhibits a dip at zero applied field. The minor loop (green) confirms the observed dip is a spin dependent effect, caused by paramagnetic moment formation. The best fit to the Hanle precession data (inset) yields $g_e^*$=5.86. d, The field dependent $T_2^total$ used to fit the Hanle data in (c).}
\end{figure}

We investigate the effect of lattice vacancy defects in graphene. Several theoretical works suggest the similarity of magnetism due to vacancies and hydrogen-doping \cite{Yazyev:2007,Soriano:2011}, as both should create magnetic moments in graphene due to the removal/hybridization of $p_z$-orbitals. It is therefore reasonable to expect that similar effects will be observable in graphene spin transport following the introduction of lattice vacancies. To induce vacancies on pristine SLG spin valves, {\it{in-situ}} Ar-sputtering is performed at a sample temperature of 15 K. Argon partial pressures of 1$\times$10$^{-6}$ torr and energies between 100 eV and 500 eV combined with short sputtering times (several seconds) produce dilute lattice vacancies. Prior to exposure to Ar-sputtering, the SLG device exhibits a $\Delta R_{NL}$ of 38.9 $\Omega$ at $V_G-V_D$=20 V and displays no dip in non-local resistance at zero applied field (Fig. \ref{fig:S5}a). Fitting of the corresponding precession data (inset of Fig. \ref{fig:S5}a) results in values of $\tau^{so}$=859 ps and $D$=0.023 m$^2$/s for the pristine SLG device. The black (red) curve presented in Fig. \ref{fig:S5}b displays $\rho_G$ before (after) sputtering. After the introduction of vacancies, the resistivity is substantially increased and the mobility is reduced from 4945 cm$^2$/Vs to 949 cm$^2$/Vs. Ar-sputtering results in a large decrease in the magnitude of $\Delta R_{NL}$ as well as the emergence of a dip in $R_{NL}$ at zero applied field (Fig. \ref{fig:S5}c). The minor loop, shown in green, indicates the observed dip is due to a decrease in the spin signal at low fields, signifying the formation of paramagnetic moments. The Hanle data (Fig. \ref{fig:S5}c inset) narrows following Ar-sputtering. The Hanle data combined with fitting the dip in $R_{NL}$ yields values of $g_e^*$=5.86, $\Delta B$=13.9 mT, $\tau_c$=64.1 ps, and the field dependent $T_2^{total}$ shown in Fig. \ref{fig:S5}d.


\section{Properties of the exchange field}  
\label{sec6}
In this section, we discuss properties of the exchange field stemming from the formation of magnetic moments with the introduction of hydrogen. Specifically, we examine the relation between the exchange field and narrowing of the Hanle curve, and we investigate the gate dependence and accuracy of $g_e^*$. 

\subsection{Exchange field and narrowing of the Hanle curve}
\label{subsec6A}
Conventional Hanle analysis, as described at the beginning of supplemental section \ref{sec4}, consists of fitting spin precession data to the Hanle equation (\ref{S14}), yielding values for the spin lifetime ($\tau^{so}$), the diffusion coefficient ($D$), and the amplitude ($S$). This relies on the assumption, $g_e^*$=2, and the absence of an exchange field. In conventional Hanle analysis, a narrowing of the Hanle curve is typically associated with an increase of the spin lifetime. Therefore, a valid question is whether the observed narrowing in the spin precession data after hydrogen doping is due to an enhanced spin lifetime instead of the emergence of an exchange field. Comprehensive analysis comprising the full data set (conductivity, non-local spin resistance ($R_{NL}$), and Hanle spin precession) provides compelling evidence that the narrowing of the Hanle curve is due primarily to an exchange field as opposed to enhanced spin lifetime. First, an increase in the spin lifetime cannot explain the observed dip in $R_{NL}$, while a fluctuating exchange field explains the dip and lineshape very well (see supplemental sections \ref{sec3} and \ref{sec4}). Second, conventional Hanle analysis ($g_e^*$=2) of the hydrogen-doped sample A yields values for the diffusion coefficient that are inconsistent with the values obtained from the conductivity (differ by a factor of $\sim$6) and values for spin lifetime $T_2^{total}$ that are inconsistent with the values obtained from the non-local spin resistance (differ by factor of 5-60). These inconsistencies can be resolved if an exchange field is present ($g_e^*\!>\!2$). Key features of the full data set emerge only after hydrogen doping and are best explained with a single effect, providing strong evidence for the presence of exchange fields. 

\begin{figure}
\includegraphics[scale=.85]{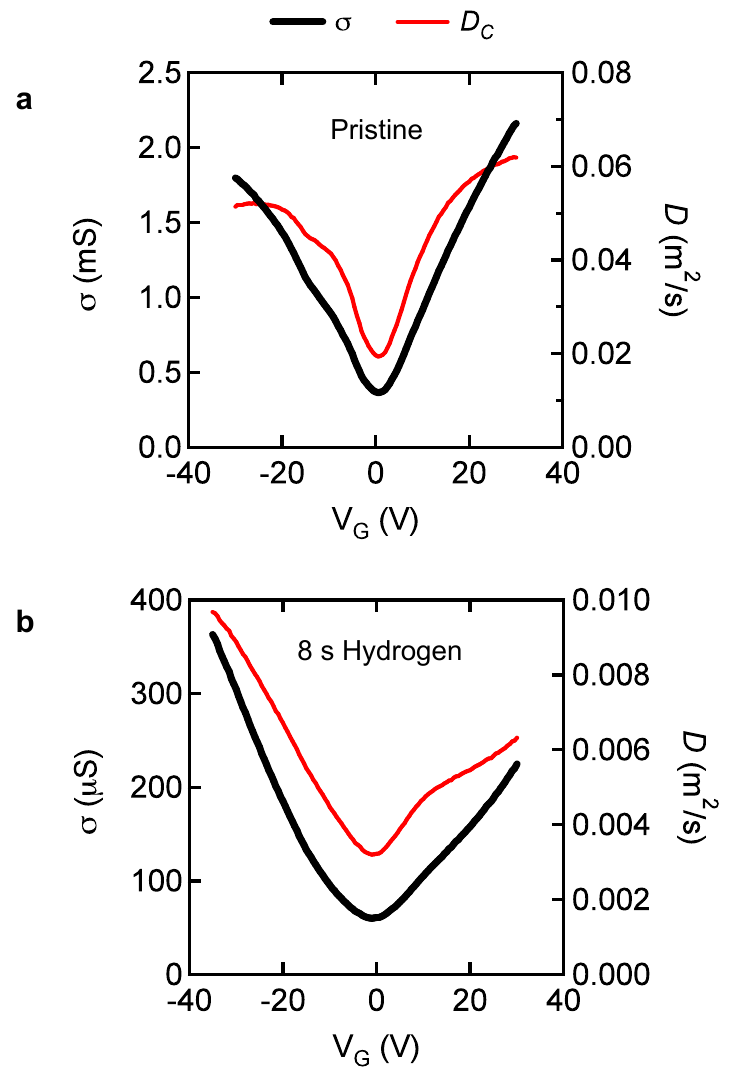}
\caption{\label{fig:S6}  Conductivity and calculated diffusion coefficient for pristine and hydrogen-doped SLG at 15 K. a, Gate dependent conductivity for pristine graphene (black curve) plotted on the left axis and the calculated diffusion coefficient, $D_C$ (red curve), plotted on the right axis. b, Gate dependent conductivity for 8 s hydrogen exposure of SLG (black curve) plotted on the left axis and $D_C$ (red curve) plotted on the right axis. }
\end{figure}

\begin{figure}
\includegraphics[scale=0.8]{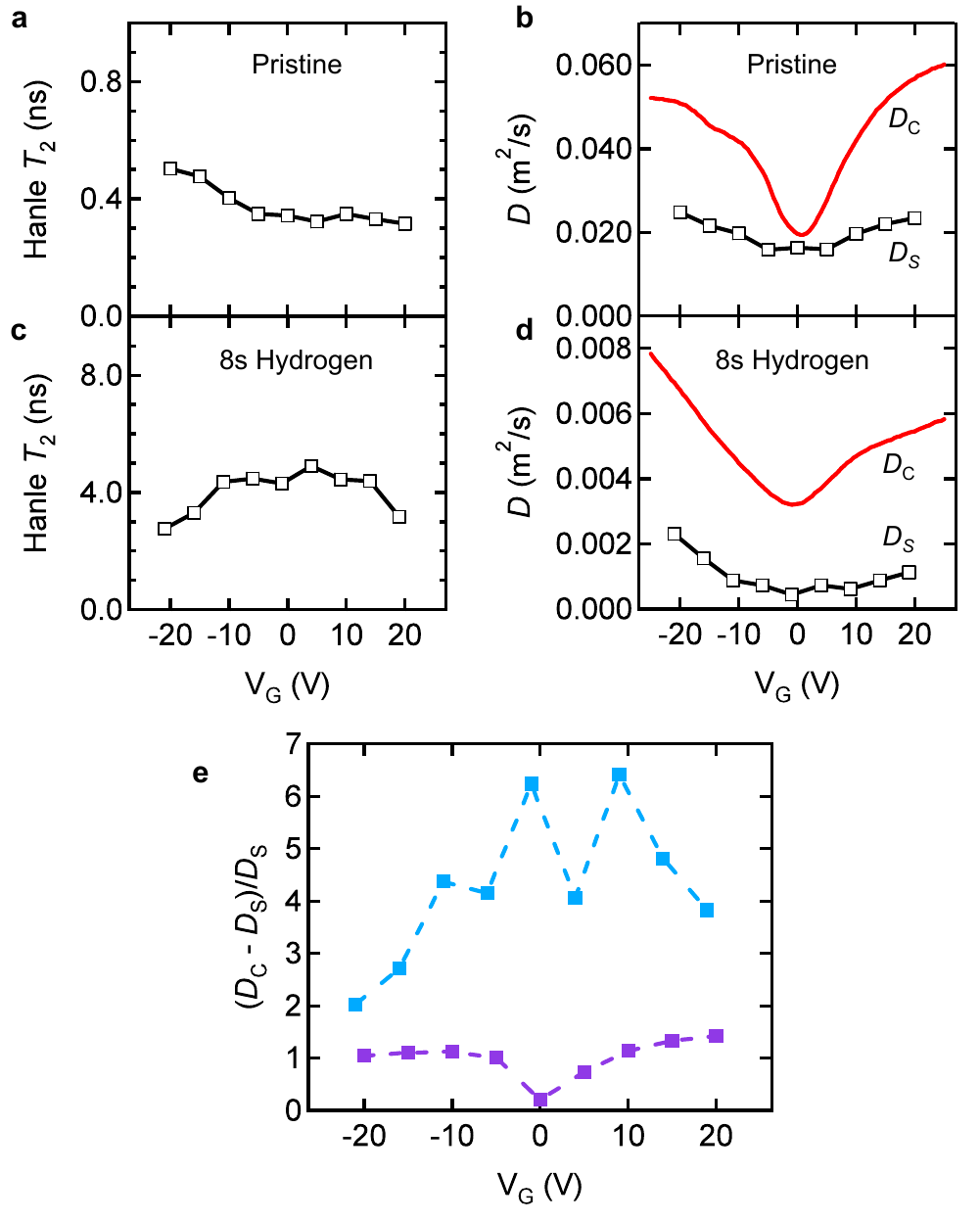}
\caption{\label{fig:S7}  Conventional Hanle fitting assuming $g_e^*$=2 (no exchange field). a, Gate dependence of the Hanle spin lifetime for pristine SLG obtained by the Hanle equation. b, Gate dependence of the diffusion coefficients $D_S$ (black open squares), obtained from the Hanle equation and the calculated diffusion coefficient, $D_C$ (red curve), obtained by calculation from the conductivity and the broadened DOS for pristine SLG. c, Gate dependence of the Hanle spin lifetime for 8 s hydrogen exposure obtained by the conventional Hanle method. d, Gate dependence of the diffusion coefficients $D_S$ (black open squares), obtained from the Hanle equation, and $D_C$ (red curve) for 8 s hydrogen-doped SLG. e, Relative difference between $D_S$ and $D_C$ for pristine (purple solid squares) and 8 s hydrogen-doped (light blue solid squares) graphene. 
}
\end{figure}

In the following, we provide a detailed analysis of the discussion outlined above. First, we investigate the Hanle spin lifetime, $T_2^{total}\!=\!\tau^{so}$, without any consideration of an exchange field (i.e. $g_e^*$=2). Fig \ref{fig:S7}a and \ref{fig:S7}c show the gate dependence of Hanle lifetimes obtained from fitting spin precession data using Hanle equation (\ref{S14}) for pristine and 8 s hydrogen exposure to sample A, respectively. As can be seen in Fig. \ref{fig:S7}c, when the Hanle fit parameters $D$ and $T_2$ are allowed to vary, best-fit values yield long spin lifetimes. Values of $D$ from the Hanle fit are denoted as $D_S$ and are displayed (black open squares) in Fig. \ref{fig:S7}b and \ref{fig:S7}d for pristine and 8 s hydrogen-doped SLG. 

Alternatively, one can use the gate-dependent conductivity to determine $D$ via the Einstein relation (\ref{S16}),
\begin{eqnarray} 
\nonumber
D=\frac{\sigma}{e^2\nu}\end
{eqnarray}
where $\sigma$ is the experimentally measured conductivity, $e$ is the electron charge, and $\nu$ is the broadened density of states (DOS). This value of diffusion coefficient is denoted as $D_C$. Spatial fluctuations of the Fermi level due to inhomogeneities in the SiO$_2$/Si substrate lead to broadening of the DOS. For Gaussian broadening \cite{Jozsa:2009}, the DOS is
\begin{equation}
\label{S29}
\nu(E) = \frac{g_v g_s 2\pi}{h^2v_F^2} \left[ \frac{2b}{\sqrt{2\pi}}e^{-\frac{E^2}{2b^2}}  + E\,\text{erf}\!\left( \frac{E}{b\sqrt{2}} \right)   \right]
\end{equation}
where $g_v$ is the valley g-factor, $g_s$ is the electron spin g-factor, $h$ is PlanckÕs constant, \mbox{$v_F$=1$\times$10$^6$ m/s} is the Fermi velocity, and $b$ is the Gaussian broadening parameter. In Figure \ref{fig:S6}a and \ref{fig:S6}b we show the conductivity (black curve) for both pristine and 8 s hydrogen exposure, respectively. These conductivity curves correspond with the resistance data of sample A as shown in Fig. 1b (main text). We find reasonable agreement for $b$=100 meV and use this throughout the remainder of this section for $D_C$. On the right axis of Figures \ref{fig:S6}a and \ref{fig:S6}b we plot the calculated diffusion coefficient (red curve) ($b$=100 meV) for pristine and 8 s hydrogen, respectively. We have found that a broadening parameter between 75 and 125 meV, which is reasonable for graphene on SiO$_2$/Si substrate \cite{Jozsa:2009,Deshpande:2009}, gives generally similar results for the present discussion. 

We next examine the difference between these two methods for determining the diffusion coefficient. Figure \ref{fig:S7}b and \ref{fig:S7}d plots $D_S$ (black open squares) and $D_C$ (red curve) for the pristine sample and the hydrogen-doped sample as a function of gate voltage, respectively. Interestingly, $D_S$ is much smaller than $D_C$ for the hydrogen-doped sample, particularly when compared to the pristine sample. To quantify this, we plot the relative difference ($D_C\!-\!D_S)\!/\!D_S$ in Figure \ref{fig:S7}e and find it to be as large as ~$\sim$6 for the hydrogen-doped sample. On the other hand, the relative difference is less than $\sim$1 for the pristine sample. Therefore, for the hydrogen-doped sample, the values of $D$ determined from the conventional Hanle method ($D_S$) and the charge transport measurement ($D_C$) are inconsistent. There are two possible explanations for the appearance of a large discrepancy in $D_S$ and $D_C$ upon hydrogen doping. First, a system with an exchange field and increased effective g-factor yields a very low value of $D_S$ when fit using conventional Hanle with $g_e^*$=2. Alternatively, it is well known that $D_S$ and $D_C$ can differ drastically if there are significant electron-electron interactions present in the system \cite{Weber:2005}. As discussed below, we find that the presence of an exchange field also resolves other inconsistencies generated by conventional Hanle fitting. 

\begin{figure}
\includegraphics[scale=0.8]{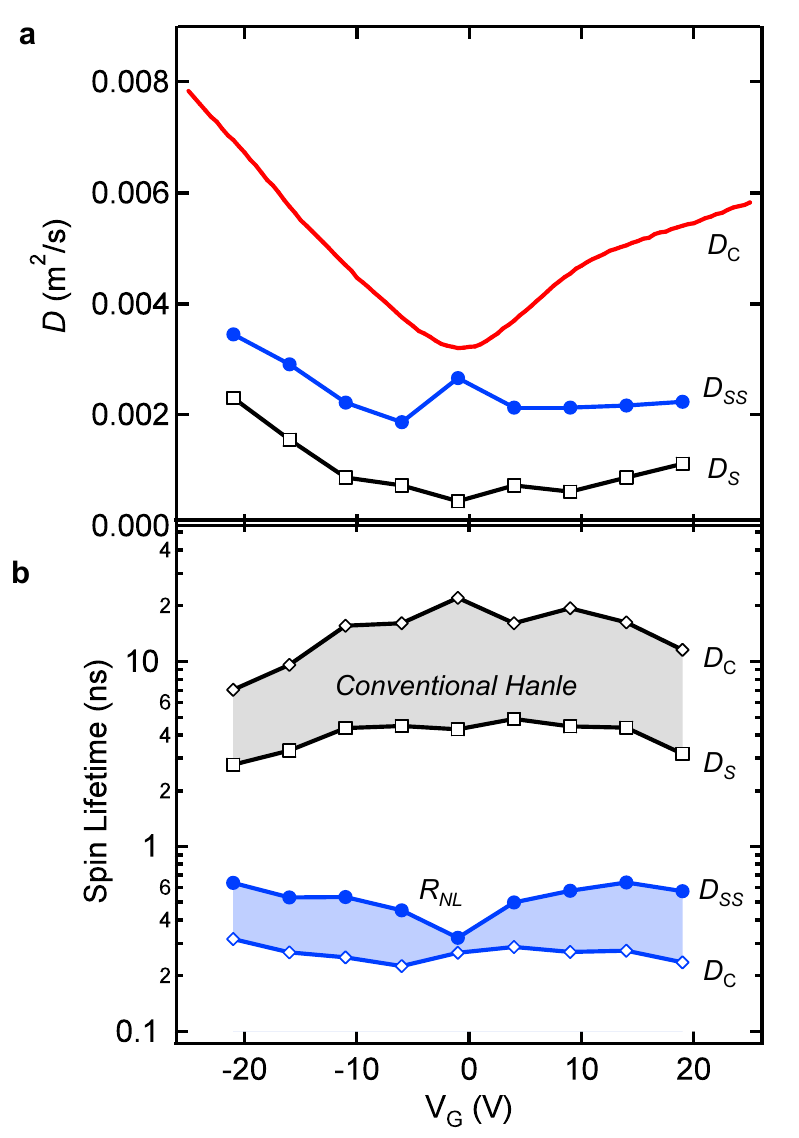}
\caption{\label{fig:S8}  Comparison of the spin lifetimes. a, Gate dependence of the diffusion coefficients $D_S$ (black open squares), $D_C$ (red curve), and $D_{SS}$ (blue solid circles) for 8 s hydrogen-doped graphene. b, Gate dependence of spin lifetimes for graphene after hydrogen exposure.  Grey shaded region represents spin lifetimes by conventional Hanle fitting with $D_C$ (black open diamonds) and $D_S$ (black open squares) without taking the possibility of magnetic moments into account ($g_e^*$=2).  A maximum $T_2$ of 33 ns is obtained at the charge neutrality point ($V_G$= -1 V) for $D_C$=0.0032 m$^2$/s. Blue shaded region represents spin lifetimes consistent with the in- plane $R_{NL}$ data. Best fits solutions for $\tau^{so}$ using $D_C$ (blue open diamonds) and $D_{SS }$ (blue solid circles) according to the procedure outlined in supplemental section \ref{sec4}. }
\end{figure}

We now consider values of spin lifetime determined by the in-plane $R_{NL}$ data and compare it to values determined from conventional Hanle fitting assuming no exchange field. Following the same procedure outlined in supplemental section \ref{sec4} from equation (\ref{S15}) to (\ref{S20}), we obtain values of $\tau^{so}$ based on best fits to the high field data of $R_{NL}$. The method utilized in section \ref{sec4} takes the diffusion coefficient, $D_{S}$, from the conventional Hanle fitting of the {\it{pristine}} sample, then scales it based on the Einstein relation according to equation (\ref{S17}). In this section, we denote this value as $D_{SS}$. The resulting spin lifetime values from $R_{NL}$ are plotted in Figure \ref{fig:S7}b as a function of gate voltage and labeled ``$D_{SS}$''. Alternatively, best-fit values for the gate dependence of spin lifetime using the calculated diffusion coefficient, $D_C$, given by equations (\ref{S16}) and (\ref{S29}), are plotted in Figure \ref{fig:S8}b (blue open diamonds) and labeled ``$D_C$''. The light blue shaded region in the range of 300-600 picoseconds represents the values of spin lifetime consistent with the non-local resistance data and is labeled ``$R_{NL}$''. We compare this with the spin lifetimes determined by conventional Hanle fitting. The spin lifetime from the conventional Hanle fitting for 8 s hydrogen exposure displayed in Figure \ref{fig:S7}c is replotted in Figure \ref{fig:S8}b and labeled ``$D_{S}$'' (black open squares). Alternative values for spin lifetime are obtained by performing the Hanle fit with the diffusion coefficient as a fixed parameter given by $D_C$ (and $g_e^*$=2). The resulting spin lifetime as a function of gate voltage is plotted in Figure \ref{fig:S8}b and labeled ``$D_C$'' (black open diamonds). The grey shaded region between 3 and 33 nanoseconds represents the values of spin lifetime consistent with the Hanle data assuming $g_e^*$=2 and is labeled ``{\it{Conventional Hanle}}''. Based on Figure \ref{fig:S8}b, the spin lifetime determined by conventional Hanle analysis (with $g_e^*$=2) is inconsistent with the spin lifetimes determined by non-local resistance.

To summarize, with the introduction of atomic hydrogen (or lattice vacancies) to SLG, conventional Hanle fitting with the assumption of $g_e^*$=2 yields two inconsistencies: (i) values of $D_S$ that are improbably low when compared to $D_C$ and (ii) values of spin lifetime that are too large compared to values obtained from the non-local resistance. Notably, both of these inconsistencies can be alleviated if $g_e^*\!>\!2$. This can be understood by considering the symmetries of the Hanle equation (\ref{S30}), 
\begin{equation}
\label{S30}
R_{NL}\!=\!S\!\int \limits_{0}^{\infty}\!\frac{e^{-L^2/4D_St}}{\sqrt{4 \pi D_S t}}\!\cos\!\left(\!\frac{g_e^* \mu_B}{\hbar}B_{app,z}t\!\right)\!e^{-t/T_2} dt
\end{equation}	
This equation is invariant under the transformation 
\mbox{$g_e^*\rightarrow c\,g_e^*$,} \mbox{$T_2 \rightarrow T_2/c$,} \mbox{$D_S\rightarrow c\,D_S,$} \mbox{$S\rightarrow c\,S$,} 
where $c$ is a constant. For a given parameter set ($g_e^*, T_2, D_S, S$), the transformed parameter set ($c\,g_e^*, T_2/c, c\,D_S, c\,S$) will generate the same Hanle curve. Therefore, if we begin with a conventional Hanle fit that assumes $g_e^*$=2, a transformation that increases $g_e^*$ (i.e. $c\!>\!1$) has the effect of decreasing $T_2$ and increasing $D_S$. This simultaneously alleviates the discrepancies in spin lifetime and diffusion coefficient mentioned earlier, and therefore provides strong evidence for the presence of exchange fields.

\begin{figure}[t]
\includegraphics[scale=0.8]{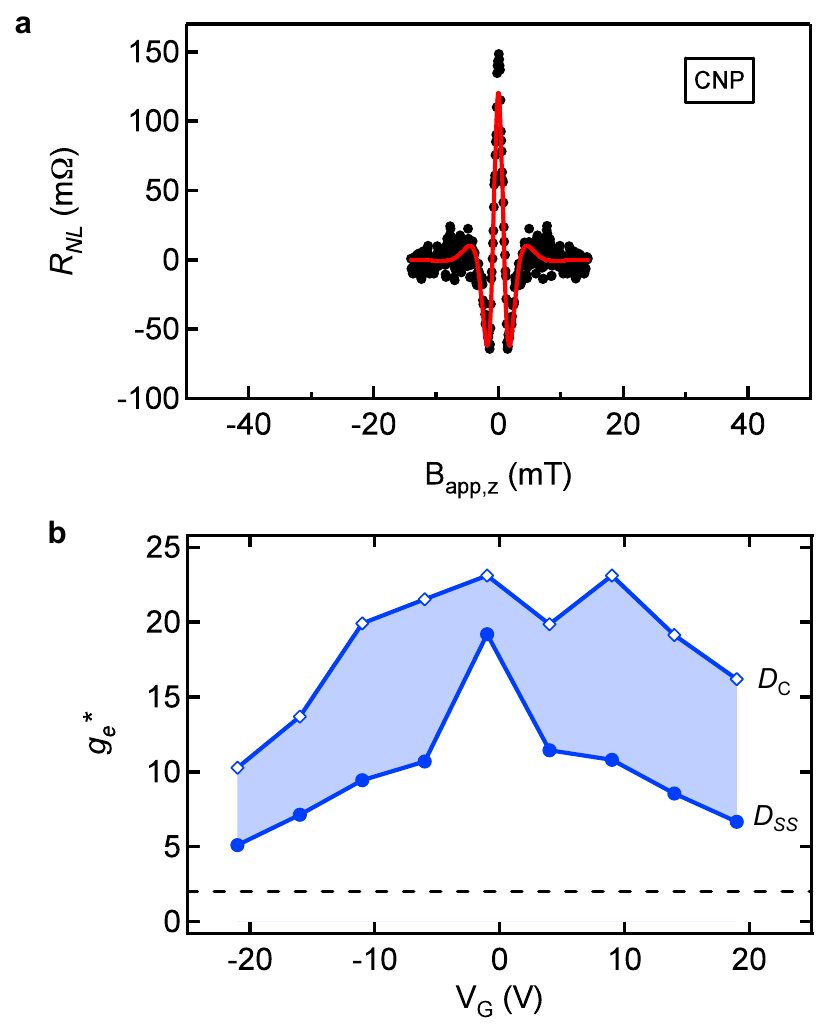}
\caption{\label{fig:S9}  Gate dependence of the enhanced precession due to the exchange field. a, Hanle precession data at the charge neutrality point ($V_G$=-1 V) for 8 s hydrogen exposure at T=15 K. A sharp increase in the observed precession due to the exchange field is seen at the charge neutrality point. Best fit to the Hanle equation (red curve) taking into account the Lorentzian shape of the spin lifetime and the scaled diffusion coefficient, $D_{SS}$, yields an effective g-factor, $g_e^*$=19.2. b, Gate dependence of $g_e^*$ for both $D_C$ (blue open diamonds) and $D_{SS}$ (blue solid circles).  }
\end{figure}

The discussion above highlights several key points about the analysis of Hanle data. In the presence of an exchange field, $g_e^*$ becomes a free parameter and the fitting parameters cannot be determined uniquely from the Hanle data alone. Therefore, it becomes necessary to analyze the in-plane $R_{NL}$ data and the Hanle data together (as detailed in supplemental section \ref{sec4}) in order to determine key parameters such as spin lifetime and $g_e^*$. The nature of this data set, with {\it{in-situ}} doping, makes it straightforward to apply this analysis, but this may not be true in other studies utilizing Hanle spin precession. Consequently, it brings to light an important question about the use of Hanle fitting in general: how does one tell whether changes in the Hanle curve are due to changes in spin lifetime or $g_e^*$? Fortunately, the above analysis leads to a useful rule of thumb: If values of $D_C$ and $D_S$ are similar for conventional Hanle fitting, this provides support for the assumption that $g_e^*$=2. This is important for future Hanle studies of spin relaxation in order to recognize when changes in spin precession data are due to changes in spin lifetime. For systems with an exchange field, this analysis motivates the need for alternative experimental techniques that can independently measure $g_e^*$ and $T_2$, such as electrically-detected electron spin resonance (ESR) and time resolved spectroscopies.

\subsection{Gate dependence and accuracy of $g_e^*$}
\label{subsec6B}
In this section we examine the gate dependence of the effective electron g-factor, $g_e^*$, due to the presence of an exchange field. Following the procedure of supplemental section \ref{sec4}, $g_e^*$ values are obtained by Hanle fits to the spin precession data for 8 s hydrogen exposure to sample A using equation \ref{S25}. Figure \ref{fig:S9}a shows the spin precession data (black closed circles) at the charge neutrality point (CNP) ($V_G$=-1 V) where the fastest precession is observed. The best fit solution (Fig. \ref{fig:S9}a red curve) to the spin precession data by equation \ref{S25} is determined through the free parameters $S$ and $g_e^*$. Fig. \ref{fig:S9}b displays the gate dependence of $g_e^*$ for both $D_{SS}$ (blue solid circles) and $D_{C}$ (blue open diamonds). The minimum value is $g_e^*$=5.1 obtained at $V_G$=-21 V and the maximum is $g_e^*$=19.2 at the CNP assuming $D_{SS}$. For $D_{C}$, the minimum $g_e^*$ value is 10.3 for $V_G$= -21 V and the maximum is $g_e^*$=21.1 at the CNP. Uncertainty of the $D$ value leads to uncertainty in $g_e^*$, again highlighting the need for techniques in graphene spintronics to directly measure $g_e^*$ and $T_2$.

\bibliography{hydrogen_supp_prl_response.bib}